\PassOptionsToPackage{table}{xcolor}
\documentclass[letterpaper,twocolumn,10pt]{article}
\usepackage{usenix2019_v3}

\usepackage{algpseudocode}
\usepackage{algorithm}
\usepackage{algorithmicx}
\usepackage{algpseudocode}
\usepackage{ifthen}

\usepackage{xcolor}
\definecolor{mycustomcolor}{rgb}{0.800, 0.400, 0.000}
\definecolor{mycommentcolor}{rgb}{0.0, 0.5, 0.0}
\usepackage{tikz}
\usepackage{listings}
\usepackage{multirow}
\usepackage{subcaption}
\usepackage{pifont}
\usepackage{tabularx}
\usepackage{xspace}
\usepackage{booktabs} 
\usepackage{adjustbox}
\usepackage{amsmath}
\usepackage{makecell}
\usepackage{url}
\usepackage{enumitem}
\usepackage{array}
\usepackage{ragged2e}

\usepackage[utf8]{inputenc} 
\usepackage[T1]{fontenc}    
\usepackage{hyperref}       
\usepackage{url}            
\usepackage{booktabs}       
\usepackage{amsfonts}       
\usepackage{nicefrac}       
\usepackage{microtype}      
\usepackage{xcolor}         
\usepackage{colortbl}
\usepackage{subcaption}
\usepackage{graphicx}
\usepackage{svg}
\usepackage{multirow}
\usepackage[table]{xcolor}
\usepackage{wrapfig}
\usepackage{threeparttable}
\usepackage{tikz}
\usepackage{tablefootnote}
\usepackage{color}
\usepackage{float}
\usepackage{diagbox}
\usepackage{tabularx}
\usepackage[strict]{changepage}
\usepackage{courier}

\newcommand{\name}{\texttt{HEC}\xspace}

\lstdefinelanguage{Graph_Representation}{
    morekeywords={combine, forcontrol, forvalue, block, load_i1, apply,arith_constant_i1,arith_xori_i1,arith_andi_i1,fanin},
    morecomment=[l]{//},
    morecomment=[s]{/*}{*/},
    morestring=[b]",
    morestring=[b]',
}

\lstdefinelanguage{Graph}{
    morekeywords={Vertex, Name, Dtype, Dimension, Block, Input, Edges, Output, Edge, Source, Vertex, Target, Vertices},
    morecomment=[l]{//},
    morecomment=[s]{/*}{*/},
    morestring=[b]",
    morestring=[b]',
    literate={\#}{{\textcolor{blue}{\#}}}1
}

\lstdefinelanguage{MLIR}{
    morekeywords={arith, affine, load, mulf, index_cast, for, store, apply, func, affine_map, return},
    morecomment=[l]{//},
    morecomment=[s]{/*}{*/},
    morestring=[b]", 
    morestring=[b]' 
}

\usepackage{inconsolata}

\newcommand{\fixme}[1]{\textcolor{red}{#1}}

\newcommand*\Circled[1]{\tikz[baseline=(char.base)]{
            \node[shape=circle,draw,inner sep=0.5pt] (char) {#1};}}

\usepackage[]{hyperref}

\usepackage{filecontents}

\begin{document}

\date{}

\title{\Large \bf HEC: Equivalence Verification Checking for Code Transformation via Equality Saturation}


\author{
{\rm Jiaqi Yin}$^{*}$
\and
{\rm Zhan Song}$^{*}$
\and
{\rm Nicolas Bohm Agostini}$^{\dagger}$
\and
{\rm Antonino Tumeo}$^{\dagger}$
\and
{\rm Cunxi Yu}$^{*}$
\and
{\rm $^{*}$\textit{University of Maryland, College Park}}
\and
{\rm $^{\dagger}$\textit{Pacific Northwest National Laboratory}}
\and
{\rm {\textit{\{jyin629,cunxiyu\}@umd.edu}}}
} 

\maketitle

\begin{abstract}
In modern computing systems, compilation employs numerous optimization techniques to enhance code performance. Source-to-source code transformations, which include control flow and datapath transformations, have been widely used in High-Level Synthesis (HLS) and compiler optimization.

While researchers actively investigate methods to improve performance with source-to-source code transformations, they often overlook the significance of verifying their correctness. Current tools cannot provide a holistic verification of these transformations. This paper introduces \name, a framework for equivalence checking that leverages the e-graph data structure to comprehensively verify functional equivalence between programs. 
\name utilizes the MLIR as its frontend and integrates MLIR into the e-graph framework. Through the combination of dynamic and static e-graph rewriting, \name facilitates the validation of comprehensive code transformations.

We demonstrate effectiveness of \name on PolyBenchC benchmarks, successfully verifying loop unrolling, tiling, and fusion transformations. \name processes over 100,000 lines of MLIR code in 40 minutes with predictable runtime scaling. Importantly, \name identified two critical compilation errors in mlir-opt: loop boundary check errors causing unintended executions during unrolling, and memory read-after-write violations in loop fusion that alter program semantics. These findings demonstrate \name practical value in detecting real-world compiler bugs and highlight the importance of formal verification in optimization pipelines.
\end{abstract}

\vspace{-3mm}
\section{Introduction}
\vspace{-3mm}




In the post-Moore era, the pressing need to enhance the runtime performance of computing systems has driven extensive research into optimization strategies.
Source-to-source transformations, such as control flow and datapath transformations, have been widely used to advance computing efficiency.
Control flow transformations optimize parameters like unrolling and tiling factors to enhance hardware parallelism and performance~\cite{agostini2022soda,wu2021ironman,sohrabizadeh2022autodse}. Furthermore, there is significant interest in datapath and gate-level transformations~\cite{yang2021equality,jia2019taso,cheng2023seer,coward2022automatic}. These combined efforts highlight a multi-faceted approach to optimize computing efficiency in various areas, including HLS~\cite{cong2022fpga,cong2011high,lai2021programming} and compiler optimization~\cite{lai2019heterocl}.
However, the stability and correctness of these transformations are often not guaranteed.
It is essential to ensure the equivalence between the programs before and after transformations.

Existing verification approaches typically focus on a single aspect of verification, thus failing to provide a comprehensive framework for verifying multi-domain transformations.
For example, Polycheck \cite{bao2016polycheck} and ISA \cite{isa} attempt to verify equivalence for affine problems, devoted to control flow transformations. 
Furthermore, Samuel et al.~\cite{coward2023datapath} have developed verification methods specifically for datapath transformations in RTL code.
These verification tools are limited in scope, and unable to offer comprehensive equivalence checking for both control flow and datapath transformations simultaneously.

As an emerging technique in compiler optimization\cite{tate2009equality, cheng2024seer}, hardware design automation\cite{ustun2022impress,woodruff2023rewriting,coward2022automatic, ustun2023equality,yin2025boole,chen2025morphic,chen2024syn}, and theorem proving\cite{detlefs2005simplify,de2008z3,bachmair1998equational,bachmair1994rewrite}, e-graph can represent multiple equivalent expressions within a single graph data structure. This capability enables possibilities of validation between optimized code against its original version. In this work, we propose \name, a formal equivalence checking framework based on e-graph rewriting.
\name is designed to provide a holistic verification of codes through equality saturation.
\name combines static rewriting rules for datapath transformations with dynamic rewriting capabilities for control flow transformations.
The main contributions of \name are summarized as follows:

\begin{figure*}[htb]
  \centering
  \begin{minipage}{0.9\columnwidth}
    \begin{lstlisting}[language=MLIR, caption=Baseline Code, label=lst:motivating1]
%av, %bv: memref<101xi1>

%true = arith.constant true
affine.for %arg1=0 to 101 {
  %1=affine.load %av[%arg1]:memref<101xi1>
  %2=affine.load %bv[%arg1]:memref<101xi1>
  %3=arith.andi %1, %2:i1
  %4=arith.xori %3, %true:i1
}
    \end{lstlisting}
  \end{minipage}
  \hspace{8mm}
  \begin{minipage}{0.9\columnwidth}
    \begin{lstlisting}[language=MLIR, caption=Variant B, label=lst:motivating2]
%av, %bv: memref<101xi1>

affine.for %arg1=0 to 101 {
  %true = arith.constant true
  %1=affine.load %av[%arg1]:memref<101xi1>
  %2=affine.load %bv[%arg1]:memref<101xi1>
  %3=arith.andi %1, %2:i1
  %4=arith.xori %3, %true:i1
}
    \end{lstlisting}
  \end{minipage}
  \hspace{8mm}
  \begin{minipage}{0.9\columnwidth}
    \begin{lstlisting}[language=MLIR, caption=Variant C, label=lst:motivating3]
%av, %bv: memref<101xi1>

%true = arith.constant true
affine.for %arg1=0 to 101 {
  %1=affine.load %av[%arg1]:memref<101xi1>
  %2=affine.load %bv[%arg1]:memref<101xi1>
  %3=arith.xori %1, %true:i1
  %4=arith.xori %2, %true:i1
  %5=arith.ori %3, %4:i1
}
    \end{lstlisting}
  \end{minipage}
  \hspace{8mm}
  \begin{minipage}{0.9\columnwidth}
    \begin{lstlisting}[language=MLIR, caption=Variant D,label=lst:motivating4]
%av, %bv: memref<101xi1>

%true = arith.constant true
affine.for %arg1=0 to 101 step 3 {
 affine.for %arg2= %arg1 to min (%arg1+3, 101) {
  %1=affine.load %av[%arg2]:memref<101xi1>
  %2=affine.load %bv[%arg2]:memref<101xi1>
  %3=arith.andi %1, %2:i1
  %4=arith.xori %3, %true:i1
 }
}
    \end{lstlisting}
  \end{minipage}
  \vspace{-4mm}
  \caption{Variants B, C, and D achieve equivalence with the baseline (Listing \ref{lst:motivating1}) through loop hoisting, De Morgan's laws, and loop tiling, respectively. This multifaceted transformation, encompassing both datapath and control flow aspects, highlights the challenges in code verification. \name offers a hybrid, combined solution to address this challenge.}
  \label{fig:motivating}
  \vspace{-4mm}
\end{figure*}

\begin{itemize}[leftmargin=*]
    \item Introducing \name, an e-graph based verification framework taking MLIR codes as inputs and performing equivalence checking. It simultaneously verifies control flow and datapath transformations, ensuring holistic equivalence checking of multi-domain transformations with equality saturation.
    \item Designing a hybrid ruleset to combine static rewriting rules with dynamic rewriting capabilities within the e-graph infrastructure, enabling the framework to adapt to runtime-dependent transformation parameters and accommodate diverse transformation patterns.
    \item Introducing a robust graph representation to model MLIR input code, encapsulating both control flow and datapath structures. This representation unifies variable renaming, loop decomposition, and operation tracking, serving as an interface for integration with the e-graph framework.
    \item Demonstrating the scalability and efficiency of \name through comprehensive evaluations on PolyBenchC benchmarks. Specifically, \name can process over 100,000 lines of MLIR code in 40 minutes. 
    \item Identifying two types of compilation errors introduced by the MLIR compiler with \texttt{mlir-opt}, in PolyBenchC \cite{pouchet2012polybench,hrishikesh2018polybenchnn} benchmark suite. 

\end{itemize}



\section{Motivating Example}









\textbf{Unified datapath/control flow verification:} Code transformation is a critical aspect of code optimization, encompassing both control flow transformations and data path transformations. Typical control flow transformations include loop unrolling, tiling, and fusion, etc~\cite{dave2019dmazerunner, yu2017advanced, ragan2013halide, tillet2019triton}. Additionally, numerous studies focus on optimizing the computation graph at the operator level for datapath optimization~\cite{jia2019taso, yang2021equality, coward2023automating, coward2022automatic}. For instance, Jia et al.~\cite{jia2019taso} optimize deep neural network (DNN) computations through graph substitution. These code optimizations underscore the necessity for rigorous verification of code transformations. In this section, we illustrate the challenges associated with code transformation verification using the example presented in Figure~\ref{fig:motivating}.

Listing~\ref{lst:motivating1} presents the baseline code that iterates over two vectors, each of length 101. For each pair of elements \texttt{a} and \texttt{b}, the loop performs the \texttt{NAND(a, b)} operation. In Listing~\ref{lst:motivating3}, a datapath transformation is demonstrated using De Morgan’s law, resulting in the \texttt{OR(a\textquotesingle, b\textquotesingle)} operation. Listing~\ref{lst:motivating2} and Listing~\ref{lst:motivating1} are functionally equivalent through loop hoisting, where the variable \texttt{\%true} is moved inside the loop body. Similarly, Listing~\ref{lst:motivating4} and Listing~\ref{lst:motivating1} are equivalent via loop tiling, where the loop on Line 4 in Listing~\ref{lst:motivating1} is split into two loops.

However, conventional verification frameworks, including PolyCheck~\cite{bao2016polycheck} and ISA~\cite{isa}, focus on only one aspect of verification. A holistic verification approach is essential to ensure the correctness and reliability of code transformations across all optimization levels, including control flow and datapath transformations. Individually verifying each transformation layer may overlook complex dependencies and interactions that can lead to subtle bugs or functional discrepancies. 

Fortunately, e-graphs~\cite{willsey2021egg} enable the simultaneous representation of multiple equivalent program expressions, effectively managing dependencies and interactions between control flow and datapath transformations. They support bidirectional rewriting rules, allowing transformations to be applied flexibly while maintaining equivalence across optimization levels. Additionally, e-graphs offer a compact representation of equivalence classes, reducing computational overhead and enhancing scalability for large codebases. These advantages collectively motivate the development of a novel e-graph based verification approach. In this paper, we introduce how the four variants in Figure \ref{fig:motivating} are integrated into the e-graph.

\textbf{Challenges:} Although e-graphs provide prominent merits for efficient verification, conventional term rewriting still poses challenges when integrating control flow transformations. For example, loop tiling in Listing~\ref{lst:motivating4} creates a new loop variable \texttt{\%arg2}, and the loop body (lines 6 to 9) replaces all instances that originally consume \texttt{\%arg1} with the newly created variable \texttt{\%arg2}. Additionally, the tiling factor (e.g., 3 in Listing~\ref{lst:motivating4}) is unknown until the compilation of the e-graph. Consequently, it is challenging to develop a static term rewriting pattern that can universally represent all tiling scenarios across diverse input codes. To effectively incorporate code equivalence checking within the e-graph framework, the rewriting rules must also include information that is only determinable at runtime, such as the specific tiling factor, variable names generated during transformation, and loop bounds that depend on input parameters. In these complex scenarios, \textit{dynamic rewriting} rules must be tailored to the specific characteristics and requirements of the input code. Static rules alone are insufficient to handle runtime-dependent factors, which can vary significantly across different codebases and transformation instances. By customizing dynamic rules, \name can accurately capture and verify intricate control flow transformation patterns, ensuring robust and reliable code equivalence checking even in highly sophisticated code structures. A more detailed discussion of the presented challenges and dynamic rewriting is provided in Section~\ref{sec:rule_generator}.

To this end, we propose \texttt{\name}, a formal equivalence verification framework utilizing e-graphs with a hybrid approach of static and dynamic rewriting. 
The integration of both static and dynamic rewriting rules within e-graph ensures robust verification of multi-domain source-to-source transformations. 
In the following sections, we demonstrate how the different variants from Listing~\ref{lst:motivating2} to Listing~\ref{lst:motivating4} are verified to be equivalent to Listing~\ref{lst:motivating1} using various approaches: \name graph representation, static rewriting, and dynamic rewriting.

\section{Background}
\label{sec:background}


\subsection{Source-to-Source Equivalence Checking}

Code equivalence checking is a critical challenge in software engineering, which involves confirming that two code segments, despite having different syntactic or structural forms, exhibit identical behaviors under all operational conditions. This verification ensures that transformations made for optimization, refactoring, or adaptation do not inadvertently alter the functional output or introduce undefined behavior.

Consider two programs, $P_A$ and $P_B$, that operate on a common input set $I$. $P_A$ produces a set of possible outputs $O_A$, while $P_B$ yields an output set $O_B$, both corresponding to the input set $I$. $P_A$ and $P_B$ are deemed functionally equivalent if and only if their output sets are identical for all possible inputs in $I$. This relationship can be formally expressed as:

\[
\forall I, \quad O_A(I) = O_B(I).
\]
This equation asserts that for every input $I$, the output $O_A(I)$ of program $P_A$ must exactly match the output $O_B(I)$ of program $P_B$, thereby affirming their functional equivalence.

Some studies attempt to establish program equivalence checking across various domains \cite{bao2016polycheck,verdoolaege2012equivalence,isa,karfa2013verification,shashidhar2005verification,pouchet2024formal}. Weilei et al.~\cite{bao2016polycheck} introduced PolyCheck, a system designed to verify the equivalence of affine programs and their transformations through dynamic analysis. However, the applicability of PolyCheck is fundamentally limited to affine transformations, thereby restricting its use in broader code transformations. Similarly, other frameworks, such as ISA~\cite{isa}, suffer from restricted transformation coverage. Generally, these frameworks are sensitive to the format of the input code, and none can comprehensively provide equivalence checking that spans data path and control flow transformations simultaneously.

\subsection{Multi-Level Intermediate Representation}

MLIR \cite{lattner2021mlir} (Multi-Level Intermediate Representation) is a compiler infrastructure within the LLVM \cite{lattner2004llvm} project. Developed to address the complexity of generating efficient code for diverse hardware targets, MLIR facilitates the co-design of high-level optimizations and hardware-specific transformations. By providing a unified interface for different levels of abstraction, from high-level algorithmic constructs down to low-level hardware instructions, MLIR supports various domain-specific optimizations. Its modular design allows developers to define custom dialects and transformations, making it highly adaptable to various computing paradigms.

Some studies focus on optimizing and transforming the MLIR infrastructure to enhance HLS and code generation. For instance, CIRCT \cite{circt} converts MLIR to RTL code, serving as an open-source HLS tool. Cheng et al. \cite{cheng2023seer} propose a MLIR based code optimization explorer based on the e-graph to address the phase-ordering problem \cite{leverett1980overview}. Tools such as HeteroCL \cite{lai2019heterocl} and ScaleHLS \cite{ye2022scalehls} transform MLIR into HLS C, while other projects like SODA \cite{agostini2022soda} and POLSCA \cite{zhao2022polsca} convert it into LLVM IR. Despite these developments, there is a lack of frameworks capable of supporting transformation verification for MLIR.

\subsection{Equality Saturation}

An e-graph (equivalence graph) is a data structure that compactly represents a congruence relation over expressions \cite{tate2009equality,nelson1980fast,nieuwenhuis2005proof}. Widely utilized in compiler optimizations, program analysis, and formal verification, e-graph efficiently manages large sets of equivalent expressions by sharing common sub-expressions, allowing it to store an exponential number of expressions in linear space.

An \textbf{e-graph} \( \mathcal{G} \) is formally defined as a tuple \( \mathcal{G} = (N, E, \phi) \):

\begin{itemize}[leftmargin=6mm]
    \item \( N \) is a finite set of nodes (also called \textbf{e-nodes}). E-nodes \( N \) represent individual expressions or sub-expressions.
    \item \( E \subseteq N \times N \) is an equivalence relation on \( N \), partitioning \( N \) into equivalence classes known as \textbf{e-classes}. E-classes are the equivalence classes induced by \( E \). Nodes within the same e-class are considered equivalent under the equivalence relation \( E \). Here \( E \) has the property of closure under congruence \cite{nelson1980fast,nelson1980techniques} which ensures that the equivalence relation is preserved under the application of operators.
    \item \( \phi: N \rightarrow \Sigma_k \times N^k \) is a labeling function that maps each node to an operator and a list of child nodes, where:
    \begin{itemize}
        \item \( \Sigma_k \) is the set of operators of operands number \( k \).
        \item \( N^k \) denotes an tuple of \( k \) child nodes from \( N \).
    \end{itemize}
\end{itemize}

 E-graphs have been used to power the optimization technique called \textbf{equality saturation}, which is an optimization process that exhaustively applies a set of rewriting rules to an initial expression within an e-graph until no new equivalence can be added. Formally, the following are defined:
\begin{itemize}
    \item \( \mathcal{G} = (N, E, \phi) \) denotes an e-graph,
    \item \( R = \{ (l_1, r_1), \dots, (l_m, r_m) \} \) is the rewriting ruleset,
    \item \( g_0 \) is the initial expression encoded in \( \mathcal{G}_0 \).
\end{itemize}

\begin{figure}[!htb]
    \centering
    \includegraphics[width=1\linewidth]{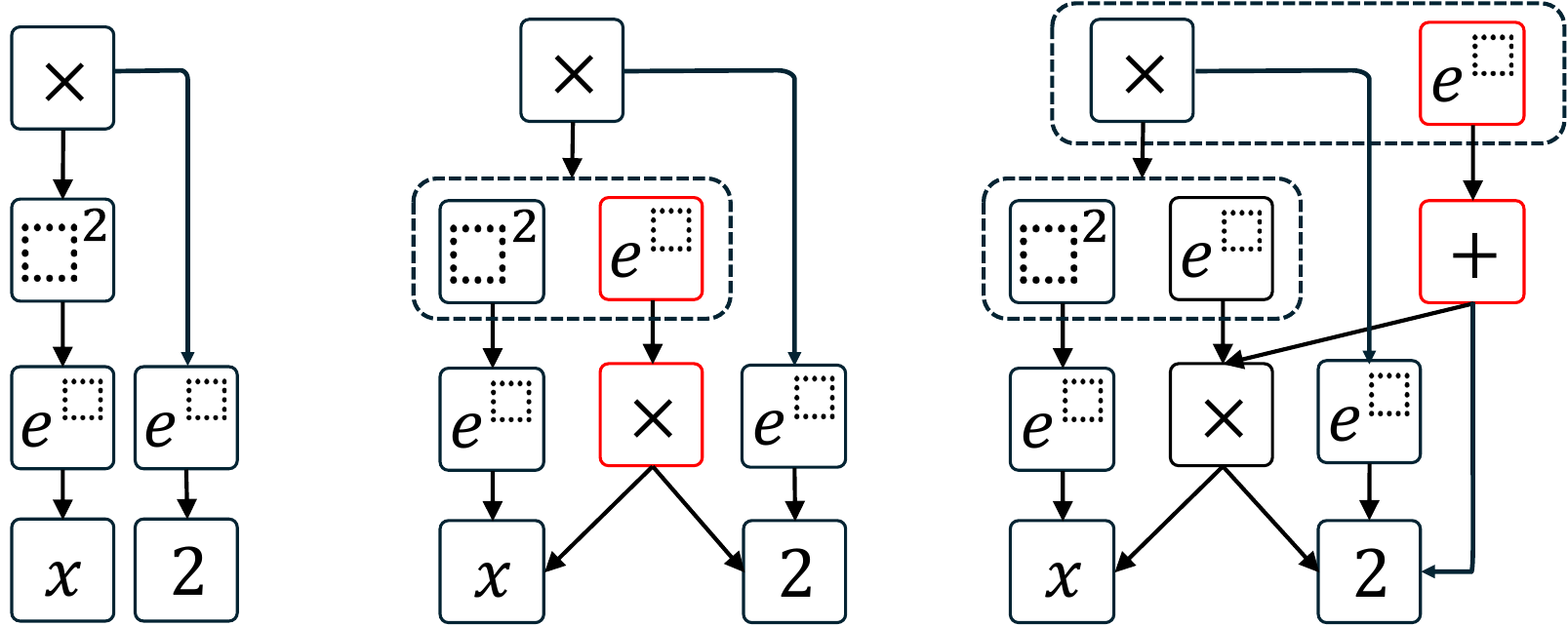}
    \caption{The e-graph rewriting for the expression ${(e^x)}^2 \times e^2$. \textbf{Left}: Initial e-graph. \textbf{Middle}: E-graph after applying the rewriting rule ${(e^x)}^2 \times e^2 \Rightarrow {e^{2x}} \times e^2$. \textbf{Right}: E-graph following the rewriting rule ${e^{2x}} \times e^2 \Rightarrow {e^{2x+2}}$. Dotted boxes represent e-classes; arrows show connections between e-nodes and their e-classes. Newly added e-nodes are highlighted in red.}
    \label{fig:e-graph}
\end{figure}

The process involves three main steps. First, construct \( \mathcal{G}_0 \) from the initial expression \( g_0 \). Second, for each pair \( (l_i, r_i) \in R \), identify matches within \( \mathcal{G}_t \) and add the equivalences \( [l_i]_E \sim [r_i]_E \) to form \( \mathcal{G}_{t+1} \) via term rewriting \cite{dershowitz2005taste}. Finally, iterate until \( \mathcal{G}_{t+1} = \mathcal{G}_t \), indicating that no further changes occur. A detailed example of e-graph exploration is shown in Figure \ref{fig:e-graph}.

Equality saturation has sparked research interests including compiler optimization \cite{tate2009equality}, hardware design automation \cite{ustun2022impress,woodruff2023rewriting,coward2022automatic}, theorem proving \cite{detlefs2005simplify,de2008z3,bachmair1998equational,bachmair1994rewrite}, and more \cite{yang2021equality,cao2023babble,vanhattum2021vectorization,panchekha2015automatically,Smith_2021}. E-graph is supported by SMT solvers, such as Z3 \cite{de2008z3,de2007efficient}, and Willsey et al. developed an advanced e-graph framework \textbf{egg} for equality saturation \cite{egglog,willsey2021egg}. In this work, \name incorporates the egg for e-graph implementation. 



Formally, the functional equivalence of two programs \( P_A \) and \( P_B \) can be defined by representing programs within a shared e-graph \( \mathcal{G} = (N, E, \phi) \) and verifying that their corresponding e-nodes \( n_A, n_B \in N \) reside within the same equivalence class. Specifically, the e-graph exhaustively applies a comprehensive set of rewriting rules \( R \) to capture all possible code transformations. If, after saturation, \( n_A \) and \( n_B \) can be united in the same e-class, then programs \( P_A \) and \( P_B \) are deemed functionally equivalent. This approach transforms program equivalence checking into determining e-class membership within the e-graph, thereby ensuring that \( P_A \) and \( P_B \) exhibit identical behaviors across all inputs \( I \).

\section{Proposed Approach}
\label{sec:approach}

\begin{figure}[!htb]
    \centering
    \includegraphics[width=1\linewidth]{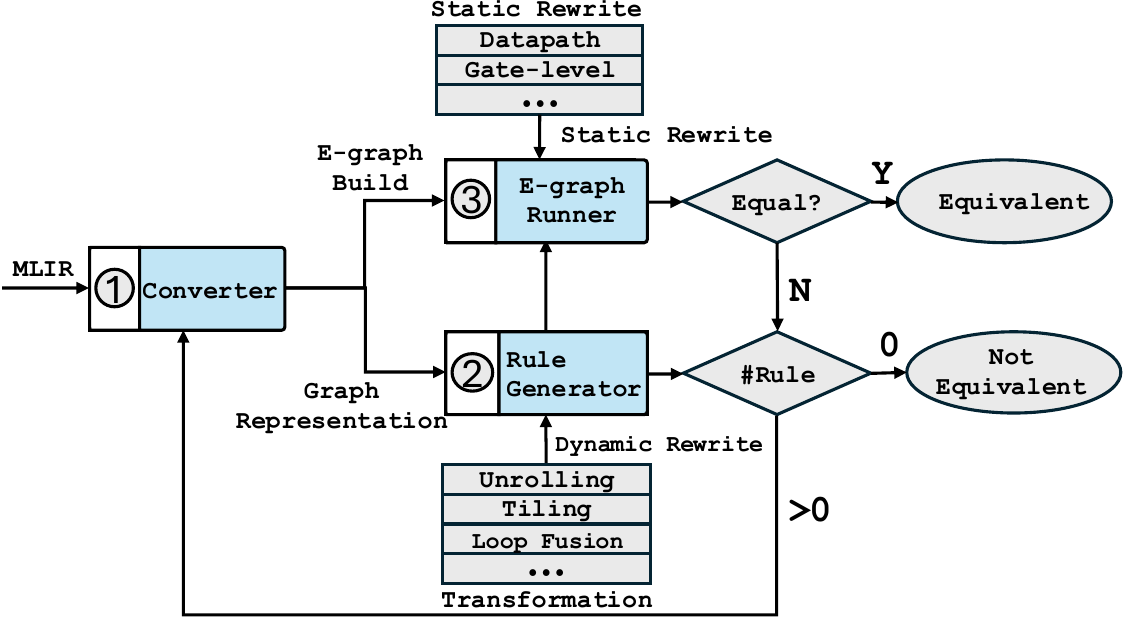}
    \caption{Overview of the \name Framework: Initially, the input MLIR code is converted into a graph-based representation through dataflow analysis (step \textcircled{1}). This graph representation serves as a foundation for constructing the e-graph. The dynamic rules (step \textcircled{2}) are specifically designed to handle complex loop transformations that are not suitable for static rewriting. Finally, the e-graph runner (step \textcircled{3}) employs a hybrid ruleset that includes both static datapath rewriting and dynamic rewriting.}
    \label{fig:overflow}
\end{figure}

Here we provide an overview of \texttt{\name}. To enhance versatility and avoid confinement to any specific programming language, \texttt{\name} utilizes MLIR as its front-end input language. Figure~\ref{fig:overflow} offers a comprehensive overview of the \texttt{\name} approach. The core functionality of \texttt{\name} lies in the verification of input code after datapath and control flow transformations, leveraging the e-graph with a hybrid static/dynamic rewriting ruleset.

\texttt{\name} accepts MLIR as input language and any framework capable of generating MLIR code can serve as the front end, such as Polygeist~\cite{moses2021polygeist} and IREE~\cite{The_IREE_Authors_IREE_2019}. \texttt{\name} incorporates a converter that processes the input MLIR code into a graph representation, acting as an interface for MLIR and the e-graph runner. Details of this graph representation are elaborated in Section~\ref{sec:graph_representation}, corresponding to step \textcircled{1} in Figure~\ref{fig:overflow}.

A major challenge within the \texttt{\name} framework is to manage control flow transformations, such as loop unrolling and tiling. Standard e-graph rewriting techniques typically struggle to accurately represent these transformations. To address this issue, we have developed a dynamic rule generator (step \textcircled{2}) that creates rewriting rules at runtime to handle control flow transformations. This enables the creation of custom rewriting rules tailored to each unique code input. The dynamic rule generation process is elaborated in Section~\ref{sec:rule_generator}.

Moreover, \texttt{\name} incorporates static datapath rewriting rules. 
In Section~\ref{sec:egraph_runner}, we present the e-graph based verification flow and the collaboration between static and dynamic rules, which are associated with step \textcircled{3} in Figure~\ref{fig:overflow}. 
Additionally, we  provide a detailed example to illustrate the process.

\vspace{-3mm}

\subsection{MLIR graph representation}
\label{sec:graph_representation}

Integrating the input MLIR code into the e-graph data structure requires translating the code into a graph representation. This translation process (step \textcircled{1}) involves two primary tasks to ensure the correctness of the mapping between the MLIR code and the graph representation. First, it needs to collect all necessary information to construct the e-graph, including dialects, operations, data type information, and other relevant metadata. Second, this involves explicitly representing loop transformation structures to ensure the graph accurately captures essential loop parameters like start, end, step, and the loop body. The generated graph representation functions as a bridge between MLIR, the e-graph runner, and the rule generator. 
Both the e-graph runner and the dynamic rule generator use the graph representation as input for further process.

\textbf{Task 1} -- The essential information encapsulated in graph representation closely mirrors the structure and characteristics of MLIR operations. We designed this representation akin to an Abstract Syntax Tree (AST), where the graph nodes correspond to individual MLIR operations. Each node comprehensively encapsulates critical information, such as the operator's name, input terms, data types, and dimensions. The interconnections between nodes are depicted as graph edges.

To construct the graph representation, we need to establish a one-to-one mapping between variable names and computation nodes. According to MLIR syntax, multiple variables may share the same variable name within their respective scopes. To address this issue, we rename all variables based on their global order of appearance in the input MLIR code. For example, if a variable \texttt{\%1} exists in multiple loop scopes, the variable in the subsequent loops will be renamed according to its appearance order. Similarly, all computation nodes are indexed based on their order of appearance. For example, the load operation in line 6 of Listing \ref{lst:motivating1} is shown in Listing \ref{lst:load_info}. 

\begin{minipage}{\linewidth}
  \begin{lstlisting}[language=Graph, caption={The graph representation attributes for the \texttt{load} operation and the term \texttt{\%2} in Listing \ref{lst:motivating1} (line 6). In this example, the \texttt{load} operation is indexed based on its order of appearance, and is assigned an index of 1
  % .\fixme{is there a reason to leave line6 empty?}
  }, label=lst:load_info]
Vertex Name: Affine_Load_1
Dtype: i1
Dimension: 1
Input Edges: %bv, %arg1
Output Edges: %2

Edge Name: %2
Source Vertex: Affine_Load_1
Target Vertices: Arith_Andi_0 
  \end{lstlisting}
\end{minipage}

\begin{figure}[!htb]
    \centering
    \includegraphics[width=0.9\linewidth]{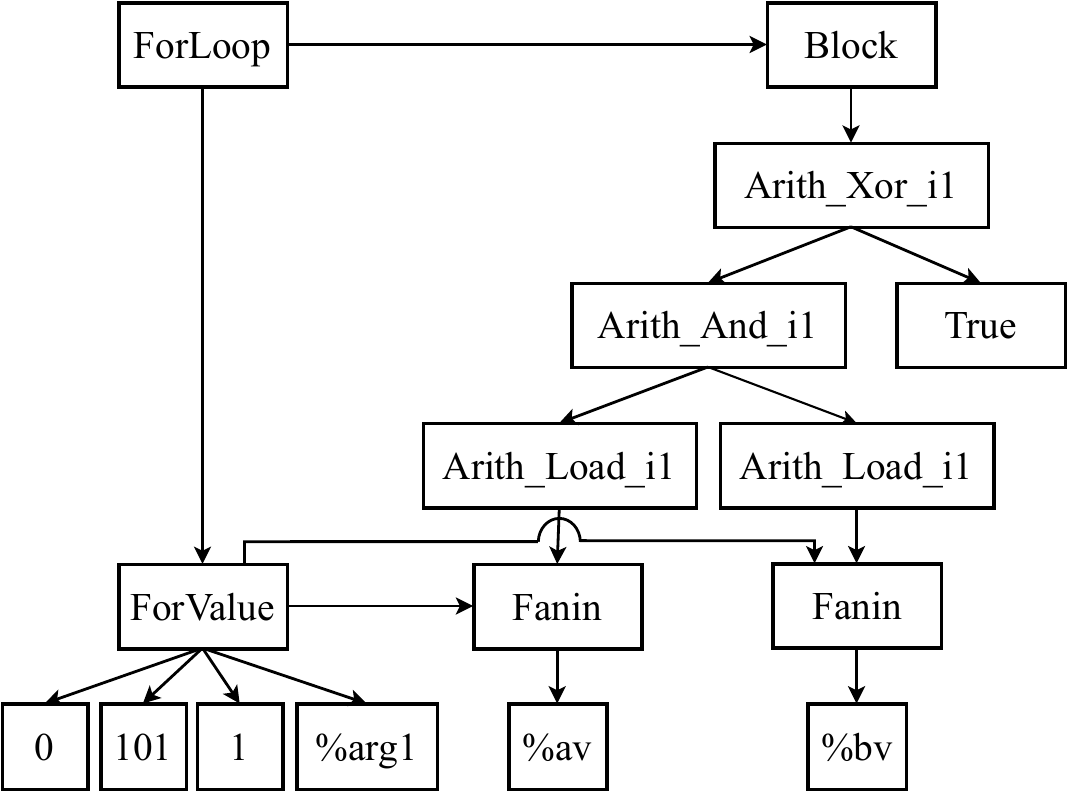}
    \caption{The dataflow graph for Listing \ref{lst:motivating1} }
    \label{fig:dataflow}
\end{figure}

\textbf{Task 2} -- The second challenge involves accurately representing \textit{for} loop constructs within our graph representation. To address this, we decompose each \textit{for} loop into two primary components: a \textit{loopvalue} component and a \textit{block} operation. A visualization of Listing \ref{lst:motivating1} is provided in Figure \ref{fig:dataflow}. Specifically, the structure of a \textit{for} loop is represented as follows:

\begin{itemize}[leftmargin=*]
    \item The \textit{loopvalue} component encapsulates the loop parameters, including the loop start value, end value, and increment step. All operations that use the loop variable as input will consume the \textit{loopvalue} component instead. 
    \item To differentiate loop variables that share the same start, end, and step values, the \textit{loopvalue} component includes the variable name as an additional input. This differentiation is not necessary for common datapath nodes such as \textit{add}. 
    \item The \textit{block} operation represents the loop body. Specifically, the inputs to the \textit{block} operation consist of the output terms from isolated operations performed within the loop body. Isolated output terms are those not consumed by any other operations within the block scope. For example, only \texttt{\%4} is isolated in Listing~\ref{lst:motivating1}. The other output terms, from \texttt{\%1} to \texttt{\%3}, are consumed by subsequent operations, thereby making \texttt{\%4} the only output for the \textit{block} operation.
    \item For operations that do not produce output data, such as \texttt{affine.store}, \name introduces a pseudo output term to maintain operation tracking. All pseudo output terms are considered isolated. 
    \item The sequence of operations within the \textit{block} operation is order-sensitive, ensuring that the original operation order is preserved within the loop block. 
    \item All hierarchical structures, including \textit{for}, \textit{func}, and \textit{if} constructs, maintain similar block operations. 
\end{itemize}

\begin{table}[!ht]
    \centering
    \small
    \renewcommand{\arraystretch}{1.2} 
    \caption{Selection of static rewriting rules integrating into \name to manage datapath transformations. All datapath rewriting rules are signage and bitwidth dependent.}
    \label{tbl:internal_rule}
    \begin{tabular}{m{1.4cm}m{2.5cm}m{3cm}}
    \hline
        Class & Left-hand Side & Right-hand Side \\ \hline
        Datapath &  \cellcolor{gray!20} $a \ll b$ & \cellcolor{gray!20} $a \times 2^b$ \\ \hline
        ~ & $(a \times b) \ll c$ & $(a \ll  c) \times b$ \\ \hline
        ~ & \cellcolor{gray!20} $a \times b \times c$ & \cellcolor{gray!20} $a \times (b \times c)$ \\ \hline
        ~ & $ (a \ll b) \ll c $ & $a \ll (b + c)$ \\ \hline
        Gate-level & \cellcolor{gray!20} $\neg(a \& b)$ & \cellcolor{gray!20} $\neg{a} \parallel \neg{b}$ \\ \hline
        ~ & $a \oplus b $ & $(a \land \neg b) \lor (\neg a \land b)$ \\ \hline
        ~ & \cellcolor{gray!20} $a \oplus 0 $ & \cellcolor{gray!20} $a$ \\ \hline
        ~ & $ \neg a $ & $a \oplus True$ \\ \hline
    \end{tabular}
\end{table}

\begin{table*}[h]
    \centering
    \footnotesize
    \caption{Selection of dynamic rewriting integrated into \name to address control flow transformations. The pattern and transformation are interchangable if they meet the conditions.}
    \label{tbl:dynamic_rule}
    \begin{tabular}{p{1.05cm}|p{3.3cm}|p{3.4cm}|p{8.3cm}}
    \hline
        \raisebox{0pt}[3ex][1.5ex]{~} & \raisebox{0pt}[3ex][1.5ex]{Left-hand Side} & \raisebox{0pt}[3ex][1.5ex]{Right-hand Side} & \raisebox{0pt}[3ex][1.5ex]{Condition} \\ \hline
        \raisebox{0pt}[7.5ex][6ex]{Unrolling} & \cellcolor{gray!20} \raisebox{0pt}[7.5ex][6ex]{\makecell[l]{ for \textbf{m1} to \textbf{n2} step \textbf{k2}: \\ \hspace*{0.6em} \texttt{Loop-body-2}}} & \cellcolor{gray!20}
        \raisebox{1pt}[7.5ex][6ex]{\makecell[l]{for \textbf{m1} to \textbf{n1} step \textbf{k1}: \\ \hspace*{0.6em} \texttt{Loop-body-1} \\ for \textbf{m2} to \textbf{n2} step \textbf{k2}: \\ \hspace*{0.6em} \texttt{Loop-body-2}}} & \cellcolor{gray!20}
        \raisebox{0pt}[7.5ex][6ex]{\parbox[t]{8.2cm}{\begin{minipage}{8.2cm}\begin{enumerate}[left=0.5em]
            \item $\lceil(\textbf{n2}-\textbf{m1})/\textbf{k2}\rceil==\lceil(\textbf{n2}-\textbf{m2})/\textbf{k2}\rceil+\lceil(\textbf{n1}-\textbf{m1})/\textbf{k1}\rceil\times(\textbf{k1}/\textbf{k2})$
            \item \texttt{Loop-body-1} is \textbf{k1}/\textbf{k2} times replication of \texttt{Loop-body-2}
        \end{enumerate}\end{minipage}}} \\ \hline
        \raisebox{0pt}[6ex][4.5ex]{Tiling} & \raisebox{0pt}[6ex][4.5ex]{\makecell[l]{for \textbf{\%1} = \textbf{m1} to \textbf{n1} step \textbf{k2}: \\ \hspace*{0.6em} \texttt{Loop-body}}} & 
        \raisebox{0pt}[6ex][4.5ex]{\makecell[l]{for \textbf{\%1} = \textbf{m1} to \textbf{n1} step \textbf{k1}: \\ \hspace*{0.6em} for \textbf{\%2} = \textbf{\%1} to \textbf{n2} step \textbf{k2}: \\ \hspace*{1.2em} \texttt{Loop-body}}} & 
        \raisebox{0pt}[6ex][4.5ex]{\parbox[t]{7.4cm}{\begin{minipage}{7.4cm} \begin{enumerate}[left=0.5em]
            \item \( \textbf{k1} == f * \textbf{k2} \) (\(f\) is the tiling factor)
            \item \( \textbf{n2} = \min(\textbf{\%1}+\textbf{k1}, \textbf{\%2}) \)
        \end{enumerate}\end{minipage}}} \\ \hline
        \raisebox{0pt}[7.5ex][6ex]{Fusion} & \cellcolor{gray!20} \raisebox{0pt}[7.5ex][6ex]{\makecell[l]{for \textbf{m} to \textbf{n} step \textbf{k1}: \\ \hspace*{0.6em} \texttt{Loop-body-1} \\ for \textbf{m} to \textbf{n} step\textbf{ k2}: \\ \hspace*{0.6em} \texttt{Loop-body-2}}} & \cellcolor{gray!20}
        \raisebox{0pt}[7.5ex][6ex]{\makecell[l]{for \textbf{m} to \textbf{n} step \textbf{k1} $\times$ \textbf{k2}: \\ \hspace*{0.6em} \texttt{Loop-body-3}}} & \cellcolor{gray!20}
        \raisebox{0pt}[7.5ex][6ex]{\parbox[t]{7.7cm}{\begin{minipage}{7.7cm}\begin{enumerate}[left=0.5em]
            \item \texttt{Loop-body-3} is \( \textbf{k2} \) times replication of \texttt{loop-body-1} plus \( \textbf{k1} \) times replication of \texttt{loop-body-2}
            \item No memory RAW violation across \texttt{Loop-body-1} and \texttt{Loop-body-2}
        \end{enumerate}\end{minipage}}} \\ \hline
        \raisebox{0pt}[7.5ex][6ex]{Coalesing} & \raisebox{0pt}[7.5ex][6ex]{\makecell[l]{for \textbf{\%1} = \textbf{m1} to \textbf{n1} step \textbf{k1}: \\ \hspace*{0.6em} for \textbf{\%2} = \textbf{\%1} to \textbf{n2} step \textbf{k2}: \\ \hspace*{1.2em} \texttt{Loop-body}}} & \makecell[l]{for \textbf{\%3} = 0 to \textbf{n1 $\times$ n2}: \\ \hspace*{0.6em} \textbf{\%1}=(floordiv \textbf{\%3} \textbf{n2}) \\ \hspace*{0.6em} \textbf{\%2}=(mod \textbf{\%3} \textbf{n2}) \\ \hspace*{0.6em} \texttt{Loop-body}} &
                \parbox[t]{7.4cm}{\begin{minipage}{7cm}\begin{enumerate}[left=0.5em]
            \item The reference of \textbf{\%1}, \textbf{\%2} is replaced by (floordiv \textbf{\%3} \textbf{n2}) and (mod \textbf{\%3} \textbf{n2}), respectively.
        \end{enumerate}\end{minipage}} \\ \hline
    \end{tabular}
\end{table*}



Another benefit of such representation is its ability to automatically unify code transformations resulting from variable movements without affecting data dependencies. For example, Listing \ref{lst:motivating2} is transformed from Listing \ref{lst:motivating1} through loop hoisting. In this transformation, the variable \texttt{\%true} is moved inside the loop body. Since \texttt{\%true} is consumed by the operation \texttt{arith.xori}, it is not isolated, and the block body does not track the term order. Both Listing \ref{lst:motivating1} and Listing \ref{lst:motivating2} correspond to the same dataflow depicted in Figure \ref{fig:dataflow}. The \name graph representation automatically unifies the loop hoisting transformation before applying e-graph rewriting.

\subsection{Hybrid E-graph Rewriting}
\label{sec:rule_generator}


In this section, we introduce the \name hybrid rewriting rule sets, which encompass both static datapath rewriting rules and dynamic rewriting rules. We specifically focus on the dynamic rewriting rules, which are designed to handle control flow transformations.

\textbf{Static Rewriting:} \name incorporates a comprehensive suite of static rewriting rules, including 62 bitwidth-dependent datapath rewriting rules. A selection of these static rewriting rules is presented in Table~\ref{tbl:internal_rule}. These static rules are specifically designed to rewrite the input program at the operator level using Boolean algebra and standard arithmetic algebra, enabling efficient datapath transformations. \


For example, as illustrated in Listing~\ref{lst:motivating3}, the framework performs \texttt{OR(a\textquotesingle, b\textquotesingle)} operations for the input operands. Here, the NOT operation $\texttt{NOT(a)}$ is expressed as $\texttt{XOR(a, True)}$ through the rewriting rule $\neg a \Leftrightarrow a \oplus \texttt{True}$. Similarly, the operations in Listing~\ref{lst:motivating1} form $\texttt{NAND(a, b)}$. These two code segments collectively constitute the left-hand side and right-hand side of the rewriting rule $\overline{a \& b} \Leftrightarrow \overline{a} \| \overline{b}$, after which the two code segments are unified into the same e-class. 




\textbf{Why we need dynamic rewriting:} Pre-defined static rewriting rules provide robust support for datapath transformations, but they face significant challenges when applied to control flow transformations. To illustrate these challenges and to highlight the need for dynamic rewriting rules, we present a simple example.


\begin{minipage}{\columnwidth}
\begin{lstlisting}[language=MLIR, caption={Unrolling variant with Listing \ref{lst:motivating1}}, label=lst:motivating5]
%av, %bv: memref<101xi1>

%true = arith.constant true
affine.for %arg1 = 0 to 100 step 2 {
%1 = affine.load %av[%arg1] : memref<101xi1>
%2 = affine.load %bv[%arg1] : memref<101xi1>
%3 = arith.andi %1, %2 : i1
%4 = arith.xori %3, %true : i1
%5 = affine.apply affine_map<(d0) -> (d0 + 1)>(%arg1)
%6 = affine.load %av[%5] : memref<101xi1>
%7 = affine.load %bv[%5] : memref<101xi1>
%8 = arith.andi %6, %7 : i1
%9 = arith.xori %8, %true : i1
}
affine.for %arg2 = 100 to 101 {
%10 = affine.load %av[%arg2] : memref<101xi1>
%11 = affine.load %bv[%arg2] : memref<101xi1>
%12 = arith.andi %10, %11 : i1
%13 = arith.xori %12, %true : i1
}
\end{lstlisting}
\end{minipage}

\begin{minipage}{\columnwidth}
\begin{lstlisting}[language=Graph_Representation,caption=Rewriting rule lhs corresponding to Listing \ref{lst:motivating1}, label=lst:lhs]
(block 
(forcontrol 
    (forvalue 0 101 1 %arg1) 
    (block 
        (arith_xori_i1 (arith_andi_i1 (load_i1 (fanin %av (forvalue 0 101 1 %arg1))) (load_i1 (fanin %bv (forvalue 0 101 1 %arg1)))) (arith_constant_i1 1))
    )))

\end{lstlisting}
\end{minipage}

\begin{minipage}{\columnwidth}
\begin{lstlisting}[language=Graph_Representation, caption=
Rewriting rule rhs corresponding to Listing \ref{lst:motivating5}, label=lst:rhs]
#map = affine_map<(d0) -> (d0 + 1)>
(combine 
(forcontrol 
    (forvalue 0 100 2 %arg1) 
    (block 
        (arith_xori_i1 (arith_andi_i1 (load_i1 (fanin %av (forvalue 0 100 2 %arg1))) (load_i1 (fanin %bv (forvalue 0 100 2 %arg1)))) (arith_constant_i1 1)) 
        (arith_xori_i1 (arith_andi_i1 (load_i1 (fanin %av (apply (map0 (forvalue 0 100 2 %arg1))))) (load_i1 (fanin %bv (apply (map0 (forvalue 0 100 2 %arg1)))))) (arith_constant_i1 1))
    )) 
(forcontrol 
    (forvalue 100 101 1 %arg2) 
    (block 
        (arith_xori_i1 (arith_andi_i1 (load_i1 (fanin %av (forvalue 100 101 1 %arg2))) (load_i1 (fanin %bv (forvalue 100 101 1 %arg2)))) (arith_constant_i1 1))
    )))
\end{lstlisting}
\end{minipage}

Consider the MLIR code shown in Listing \ref{lst:motivating5}, which is an unrolled version of Listing \ref{lst:motivating1} with an unrolling factor of 2. To establish the equivalence between these two codes, the e-graph runner requires a rewriting rule of the form \( lhs \Leftrightarrow rhs \), where the left-hand side (lhs) and right-hand side (rhs) are provided in Listings \ref{lst:lhs} and \ref{lst:rhs}, respectively. However, constructing static rewriting rules to capture such transformations presents several challenges: (1) The number of operations within the \textit{block} operation (line 5 in Listing~\ref{lst:rhs}) varies according to the unrolling factor. Static e-graph rewriting rules necessitate a predetermined number of input parameters prior to e-graph runner compilation, making it difficult to accommodate varying operation counts; (2) The isolated dataflow structures need to be isomorphic (line 5 in Listing \ref{lst:lhs} compared to lines 6, 7, and 12 in Listing \ref{lst:rhs}). Assessing the isomorphism of these dataflow structures is challenging simply using static rewriting rules; (3) Loop unrolling introduces new variables, and their metadata (such as variable names and data types) is collected at runtime. For example, the loop variable \%arg1 is created during the unrolling process in Listing~\ref{lst:motivating5}. Since this runtime-collected information must be determined prior to compilation, creating static "unrolling" e-graph rewriting rules that accommodate all unrolling scenarios becomes impractical.
In complex scenarios like loop unrolling, where the unrolling factor and variable metadata can vary dynamically based on input code, static rewriting rules prove insufficient. Therefore, dynamic rewriting rules must be customized to handle such variability.
These challenges collectively prompt us to develop dynamic rewriting rules, generated at runtime, to effectively handle complex control flow transformations.


\textbf{Dynamic Rewriting:} To overcome the limitations of static rewriting, \name generates dynamic rewriting rules at runtime, tailoring them to each unique code input to achieve high adaptability across diverse scenarios. In Step~\Circled{2}, the dynamic rule generator leverages the pre-defined transformation patterns to assess computation nodes for applicable transformations. Once suitable nodes are identified, \name creates new dynamic rewriting rules based on the graph representation.

\begin{figure}
  \begin{minipage}{\columnwidth}
  \begin{lstlisting}[language=MLIR]
affine.for %arg1 = m1 to n2 step k2:
  // Loop-body-2
  \end{lstlisting}
  \vspace{-8mm}
  \end{minipage}
  \begin{minipage}{\columnwidth}
    \centering
    \includegraphics[width=1\linewidth]{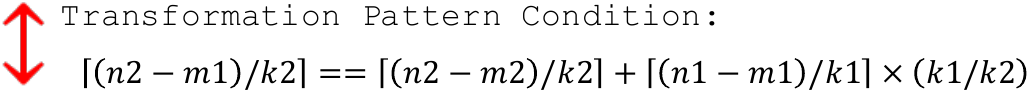}
  \vspace{-5mm}
  \end{minipage}
  \begin{minipage}{\columnwidth}
    \begin{lstlisting}[language=MLIR]
affine.for %arg1 = m1 to n2 step k2:
  // Loop-body-2
    \end{lstlisting}
  \end{minipage}
  \vspace{-9mm}
  \caption{Loop unrolling transformation pattern: the code segments at the top and bottom can be interchanged if they satisfy the required conditions in the middle.}
  \label{fig:rule}
  \vspace{-6mm}
\end{figure}

Using loop unrolling transformations as an example to demonstrate our verification approach, we establish the transformation criteria shown in Figure \ref{fig:rule}, which are used to determine whether a code segment qualifies as an unrolled variant. 
These criteria define the requirements for \textit{for} loops to be recognized as equivalent to loop unrolling transformations. 
For instance, the loops in lines 4 and 15 of Listing~\ref{lst:motivating5} satisfy the conditions outlined in Figure~\ref{fig:rule}, indicating their eligibility for the loop unrolling transformation pattern and confirming their equivalence to those in Listing~\ref{lst:motivating1}.
Additionally, we introduce a leading operator, \textit{combine}, to bind the two loop operations (line 2 in Listing~\ref{lst:rhs}). Here, we assume that all integrated rewriting rules and transformation patterns are correct. Whether the input code meets the transformation pattern conditions is automatically verified using the Z3 SMT solver~\cite{de2008z3,de2007efficient}.

\name analyzes the graph representation illustrated in Section~\ref{sec:graph_representation} to assess the eligibility of each \texttt{for} loop. For each eligible loop, \name constructs a dynamic rule \( lhs \Leftrightarrow rhs \). For instance, Listing~\ref{lst:lhs} and Listing~\ref{lst:rhs} show the \( lhs \) and \( rhs \) of dynamic rewriting rules in Listing~\ref{lst:motivating5} unrolling, respectively. A selection of dynamic rewriting patterns is shown in Table \ref{tbl:dynamic_rule}.

Similarly, Listing~\ref{lst:motivating4} is verified to be equivalent to Listing~\ref{lst:motivating1} through dynamic rewriting. Specifically, the \textit{for} loop in line 4 of Listing~\ref{lst:motivating4} is identified as an eligible computation node for loop tiling, and customized dynamic rewriting rules for tiling are established to achieve equivalence with Listing~\ref{lst:motivating1}.

\textbf{Extensibility:} \name supports extensibility through a systematic approach for incorporating new transformation patterns. For control flow transformations, users must formalize the transformation pattern along with its correctness conditions (similar to those in Table \ref{tbl:dynamic_rule}), while datapath transformations require the corresponding static rewriting rules. To ensure soundness, \name employs a dual validation strategy: static datapath transformation rules are derived from mathematically proven algebraic identities such as De Morgan's laws and arithmetic associativity, making them sound by construction, while dynamic rules for control flow transformations undergo formal verification using the Z3 SMT solver to ensure mathematical correctness. For instance, the loop unrolling condition specified in Table \ref{tbl:dynamic_rule} is mathematically verified to guarantee preservation of the iteration space, thereby maintaining semantic equivalence between the original and transformed code.

\name limits Z3 usage to verifying dynamic rule pattern conditions rather than incorporating it into entire saturation process, thereby avoiding scalability bottlenecks. Our testing confirms that the dynamic rule generation, including Z3 checks, typically completes in under one second, a negligible fraction of the overall verification runtime, which is predominantly consumed by e-graph saturation.

\subsection{\name Verification Runner}

\begin{figure}[!htb]
  \begin{minipage}{\columnwidth}
    \begin{lstlisting}[language=MLIR,label=lst:motivating_baseline]
%0 = arith.index_cast %arg0 : i32 to index
affine.for %arg2 = 0 to %0
  %1 = affine.load %arg1[%arg2] : memref<?xf64>
    \end{lstlisting}
    \vspace{-8mm}
  \end{minipage}
  \begin{minipage}{\columnwidth}
    \begin{lstlisting}[language=MLIR,label=lst:motivating_nested_unroll]
#map1 = affine_map<()[s0] -> (((s0 floordiv 2) floordiv 3) * 6)>
#map2 = affine_map<()[s0] -> ((s0 floordiv 2) * 2)>
#map3 = affine_map<()[s0] -> ((s0 floordiv 2) * 2 + ((s0 mod 2) floordiv 3) * 3)>

affine.for %arg2 = 0 to #map1()[%0] step 6
  %1 = affine.load %arg1[%arg2] : memref<?xf64>
  %2 = affine.apply affine_map<(d0) -> (d0 + 1)>(%arg2)
  %3 = affine.load %arg1[%2] : memref<?xf64>
  %4 = affine.apply affine_map<(d0) -> (d0 + 2)>(%arg2)
  %5 = affine.load %arg1[%4] : memref<?xf64>
  %6 = affine.apply affine_map<(d0) -> (d0 + 1)>(%4)
  %7 = affine.load %arg1[%6] : memref<?xf64>
  %8 = affine.apply affine_map<(d0) -> (d0 + 4)>(%arg2)
  %9 = affine.load %arg1[%8] : memref<?xf64>
  %10 = affine.apply affine_map<(d0) -> (d0 + 1)>(%8)
  %11 = affine.load %arg1[%10] : memref<?xf64>
affine.for %arg2 = #map1()[%0] to #map2()[%0] step 2
  %1 = affine.load %arg1[%arg2] : memref<?xf64>
  %2 = affine.apply affine_map<(d0) -> (d0 + 1)>(%arg2)
  %3 = affine.load %arg1[%2] : memref<?xf64>
affine.for %arg2 = #map2()[%0] to #map3()[%0] step 3
  %1 = affine.load %arg1[%arg2] : memref<?xf64>
  %2 = affine.apply affine_map<(d0) -> (d0 + 1)>(%arg2)
  %3 = affine.load %arg1[%2] : memref<?xf64>
  %4 = affine.apply affine_map<(d0) -> (d0 + 2)>(%arg2)
  %5 = affine.load %arg1[%4] : memref<?xf64>
affine.for %arg2 = #map3()[%0] to %0
  %1 = affine.load %arg1[%arg2] : memref<?xf64>

    \end{lstlisting}
  \end{minipage}
  \vspace{-7mm}
  \caption{MLIR Nested Unrolling Case Study: The bottom MLIR code is produced by applying nested unrolling with factors of 2 and 3 to the top MLIR code.}
  \label{fig:nested_unrolling}
\end{figure}

In this section, we will elaborate more details about e-graph verification runner and use a specific example to illustrate the verification process.

\begin{figure*}[htb]
  \centering
  \begin{minipage}{1\textwidth}
    \centering
    \subcaptionbox{In the first iteration, \name inserted pseudo \textit{Combine} nodes to bind leaf $loop_1/loop_2$ and $loop_3/loop_4$. Then, the original sub-expression, indicated by a dotted line, is transformed into a new expression, represented by a solid line. Following this, the e-graph runner implements dynamic rules for loop\textsubscript{1}/loop\textsubscript{2} and loop\textsubscript{3}/loop\textsubscript{4} (highlighted by light blue nodes), merging them into loop\textsubscript{12} and loop\textsubscript{34} (illustrated with dark red nodes). However, \name cannot verify the equivalence in the first iteration.
      \label{fig:egraph_1}}
      {\includegraphics[width=1\linewidth]{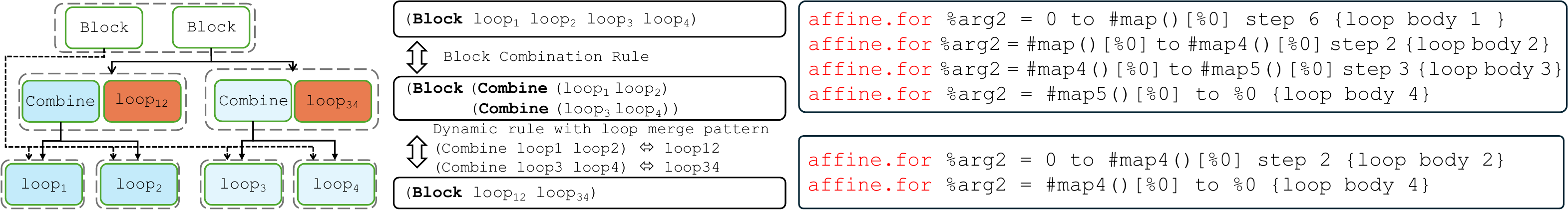}}
    \subcaptionbox{In the second iteration, the rule generator introduces a new rule, merging loop\textsubscript{12} and loop\textsubscript{34} into loop\textsubscript{1234}, similar to the procedure in the initial iteration. Subsequently, the e-graph runner validates the equivalence between the two input MLIR codes.
      \label{fig:egraph_2}}
      {\includegraphics[width=1\linewidth]{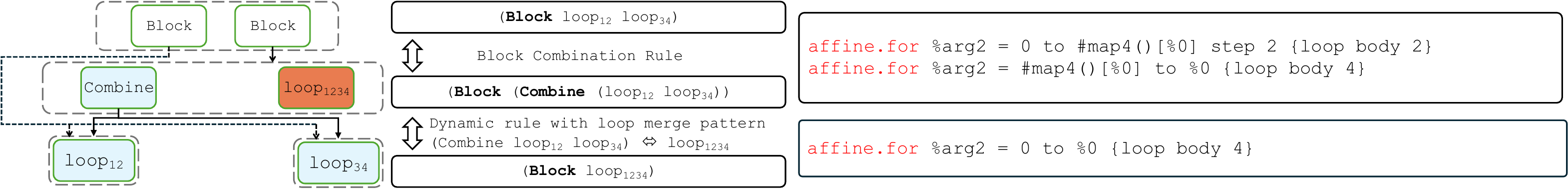}}
  \end{minipage}
  \caption{E-graph verification for the motivating example provided in Figure \ref{fig:nested_unrolling}. The e-graph operates through two iterations, with the rule generator introducing dynamic rules to the e-graph runner during each iteration. Each \textit{loop} operator represents a \textit{for} loop structure, as depicted in Figure \ref{fig:dataflow}, for brevity purposes.}
  \label{fig:egraph}
\end{figure*}

\textbf{Initial E-graph Construction}: \name takes the graph representation as input. It gathers functional information and constructs the corresponding e-graph. The algorithm for e-graph construction is detailed in Algorithm \ref{alg:egraph_construction}. Before construction, it builds the topological order for each operation based on the dependence relationship. Subsequently, \name incrementally inserts operation nodes into the e-graph following the topological order of the graph nodes. Specifically, operations are inserted into the e-graph from the leaf nodes to the root nodes to ensure that child nodes are processed beforehand. Each insertion generates an identifier representing the e-nodes and connects these e-nodes with their respective input identifier.

\algnewcommand{\Input}[1]{\Require{#1}}
\algnewcommand{\Output}[1]{\Ensure{#1}}
\algrenewcommand{\algorithmicrequire}{\textbf{Input:}}
\algrenewcommand{\algorithmicensure}{\textbf{Output:}}

\begin{algorithm}
\caption{E-graph Construction}
\scriptsize
\label{alg:egraph_construction}
\begin{algorithmic}[1]
   \Input{ Vertex list $V$ from parsing netlist} 
   \Output:{ E-graph $G_e$}

    \State Initialize $vmap$ as empty HashMap
    \For{each node $v$ in TopoSort($V$)} \Comment{From leaf to root}
    \State $in\_id \leftarrow \lbrack \rbrack$ 
    \For{each input $i$ for node $v$}
    \State $in\_id$.pushback($vmap$[$i$]) \Comment{All children are inserted}
    \EndFor
    \State $id \leftarrow $ $G_e$.insert($v, in\_id$) \Comment{Insert $v$ to e-graph}
    \State $vmap[v] = id$
    \EndFor
\end{algorithmic}
\end{algorithm}

\textbf{\name Verification}: To demonstrate the verification procedure, we use the example shown in Figure \ref{fig:nested_unrolling}. The bottom MLIR code presents a nested loop unrolled with factors of 2 and 3, derived from the top MLIR code. The verification process is depicted in Figure \ref{fig:egraph}. For simplicity, the \textit{for} loop structures in Figure \ref{fig:dataflow} are simplified to single-loop operations.

Initially, the ruleset includes only the predefined static rules without any dynamic rules. If the differences between the two MLIR codes fall within the scope addressable by static rewriting rules, the e-graph runner reports the codes as functionally equivalent. However, if the MLIR codes involve control flow transformations, the e-graph runner cannot affirm equivalence. In such cases, the graph representation is forwarded to the rule generator to apply the control flow transformation patterns.

Figure \ref{fig:egraph_1} illustrates the e-graph exploration during the initial iteration. The rule generator analyzes the graph representation and creates dynamic rules for candidate operations. \name then creates a pseudo \textit{combine} node to bind candidates \textit{for} loop operations. The \textit{combine} nodes are created automatically for each pair of candidate nodes of dynamic rules. Finally, \name unifies the e-classes of \textit{combine} node and the fused loop operations using dynamic rewriting rules. For example, the pairs of \textit{for} loop structures from lines 5–17 and lines 21–27 in the unrolled MLIR code of Figure \ref{fig:nested_unrolling} (indicated by dotted lines in Figure \ref{fig:egraph_1}) are eligible for the loop unrolling transformation pattern. The e-graph runner creates a pseudo \textit{combine} node as their parent e-node and builds dynamic rules to merge the \textit{combine} node with \textit{loop}\(_{12}\) into the same e-class (indicated by solid lines). Consequently, \textit{loop}\(_{1}\)/\textit{loop}\(_{2}\) and \textit{loop}\(_{3}\)/\textit{loop}\(_{4}\) are merged into two loops named \textit{loop}\(_{12}\) and \textit{loop}\(_{34}\), respectively. The MLIR codes for both the original loops and the unrolled version are displayed on the right side of Figure \ref{fig:egraph_1}.

However, the e-graph runner is still unable to ascertain the equivalence between the two input codes. An additional iteration is necessary for confirmation. Similar to the approach in the first iteration, the rule generator proposes a new dynamic rule for merging \textit{loop}\(_{12}\) with \textit{loop}\(_{34}\) into a single \textit{loop}\(_{1234}\). The illustration of the second iteration is shown in Figure \ref{fig:egraph_2}. Consequently, the two MLIR codes in Figure \ref{fig:nested_unrolling} are unified into the same e-class. It is worth noting that Figure \ref{fig:egraph_2} shows only one possible verification path for clarity. In practice, the e-graph simultaneously explores multiple combination patterns, such as merging \textit{loop}\(_{2}\) and \textit{loop}\(_{3}\) into \textit{loop}\(_{23}\), which could lead to alternative paths like \textit{loop}\(_{1}\), \textit{loop}\(_{2}\), \textit{loop}\(_{3}\), \textit{loop}\(_{4}\) $\rightarrow$ \textit{loop}\(_{1}\), \textit{loop}\(_{23}\), \textit{loop}\(_{4}\) $\rightarrow$ \textit{loop}\(_{123}\), \textit{loop}\(_{4}\) $\rightarrow$ \textit{loop}\(_{1234}\). The e-graph's congruence closure property ensures that all valid transformation sequences leading to equivalent programs are automatically discovered and unified within the same e-class. If the e-graph runner still cannot determine the equivalence and the rule generator fails to create new dynamic rules for further evaluation, \name will conclude that the input MLIR codes are not equivalent. It is important to note that both the e-graph verification runner and the rule generator utilize the graph representation as their input. During each iteration, \name is equipped with an inverter, which converts the e-graph back into the graph representation to facilitate further processing in the subsequent iteration. Besides, the verification flow is fully automated. \name offers an interface for adding control flow transformation patterns, facilitating broader equivalence verification through an expanded set of rewriting rules and verification patterns.

\textbf{Completeness and Soundness}: \name, like any verification system based on rewriting rules, is inherently incomplete with respect to the universe of all possible code transformations. If an equivalence exists but requires a transformation rule not included in our framework, or if the saturation process is terminated due to computational limits before full exploration is complete, \name may fail to establish equivalence even when it exists, resulting in false negatives (flagging equivalent programs as nonequivalent).

\name guarantees soundness through a dual validation mechanism that ensures it will never produce false positives (declaring two programs equivalent when they are not). For static datapath transformation rules, soundness is achieved by deriving rules from mathematically proven algebraic identities such as De Morgan's laws and arithmetic associativity, making them sound by construction. For dynamic control flow transformation rules, pattern conditions are formally verified using the Z3 SMT solver to ensure mathematical correctness.

\label{sec:egraph_runner}

\section{Evaluation}
\label{sec:evaluation}

\begin{figure*}[htbp]
    \centering
    \begin{subfigure}[b]{0.16\textwidth}
        \includegraphics[width=\textwidth]{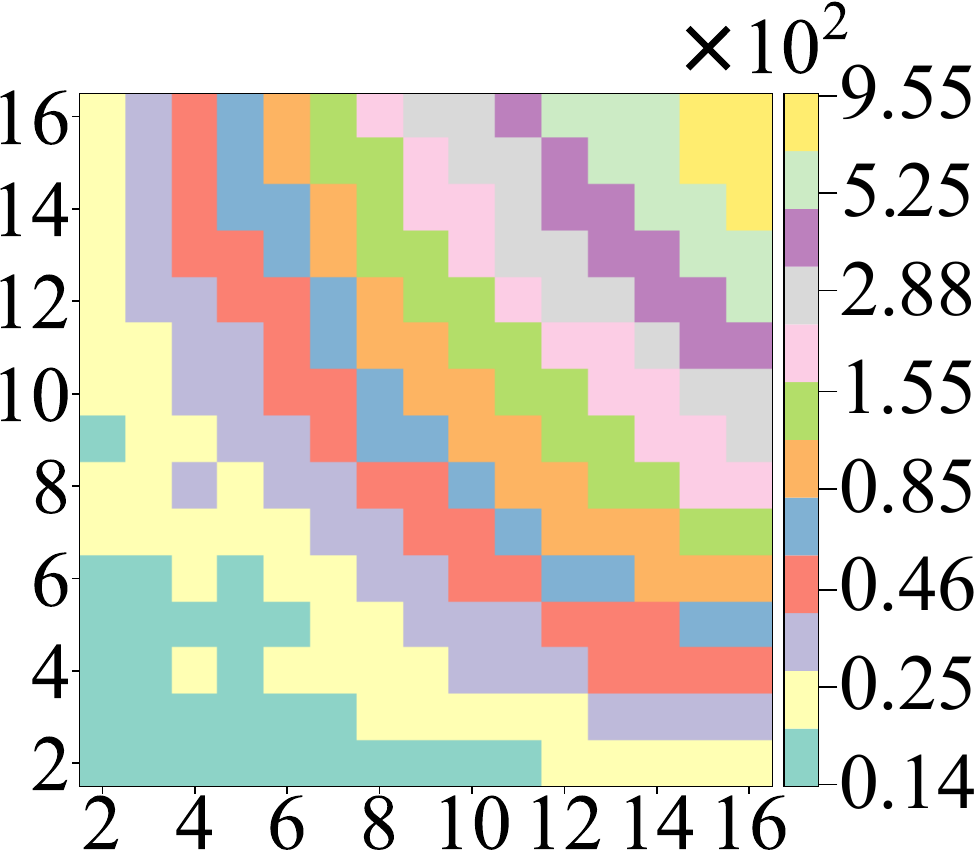}
        \caption{2mm}
        \label{fig:heatmap:2mm}
    \end{subfigure}
    \hfill
    \begin{subfigure}[b]{0.16\textwidth}
        \includegraphics[width=\textwidth]{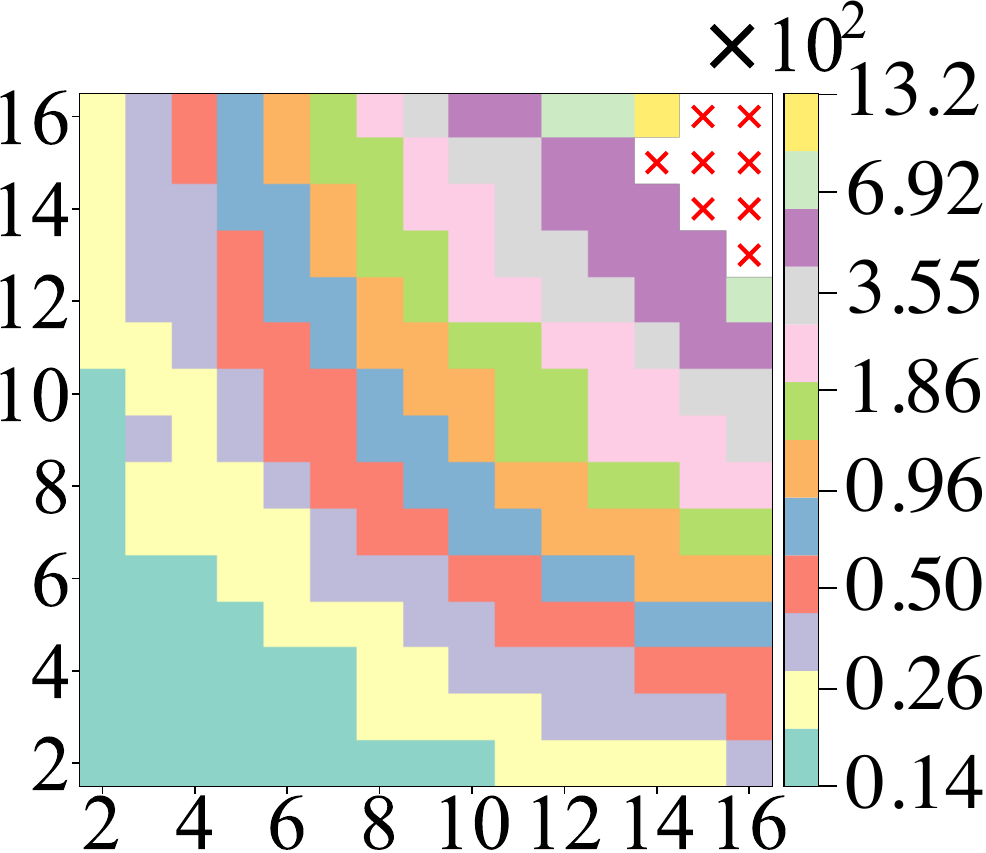}
        \caption{Jacobi\_1d}
        \label{fig:heatmap:jacobi_1d}
    \end{subfigure}
    \hfill
    \begin{subfigure}[b]{0.16\textwidth}
        \includegraphics[width=\textwidth]{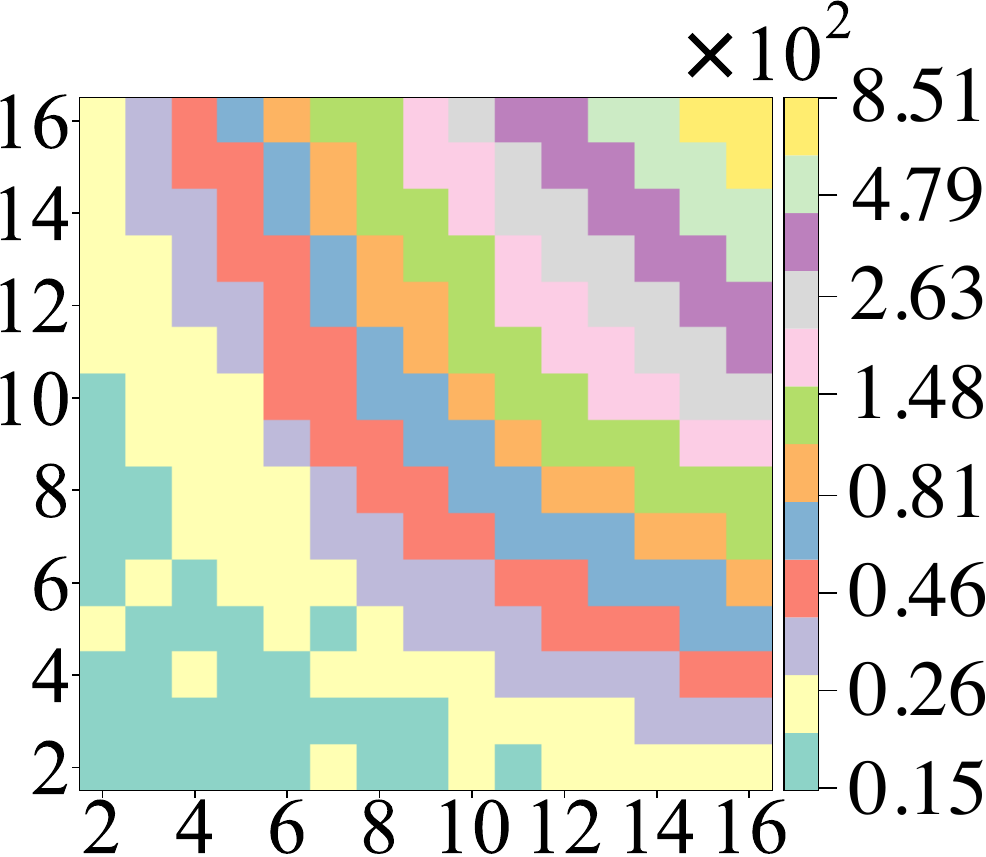}
        \caption{Lu}
        \label{fig:heatmap:lu}
    \end{subfigure}
    \hfill
    \begin{subfigure}[b]{0.16\textwidth}
        \includegraphics[width=\textwidth]{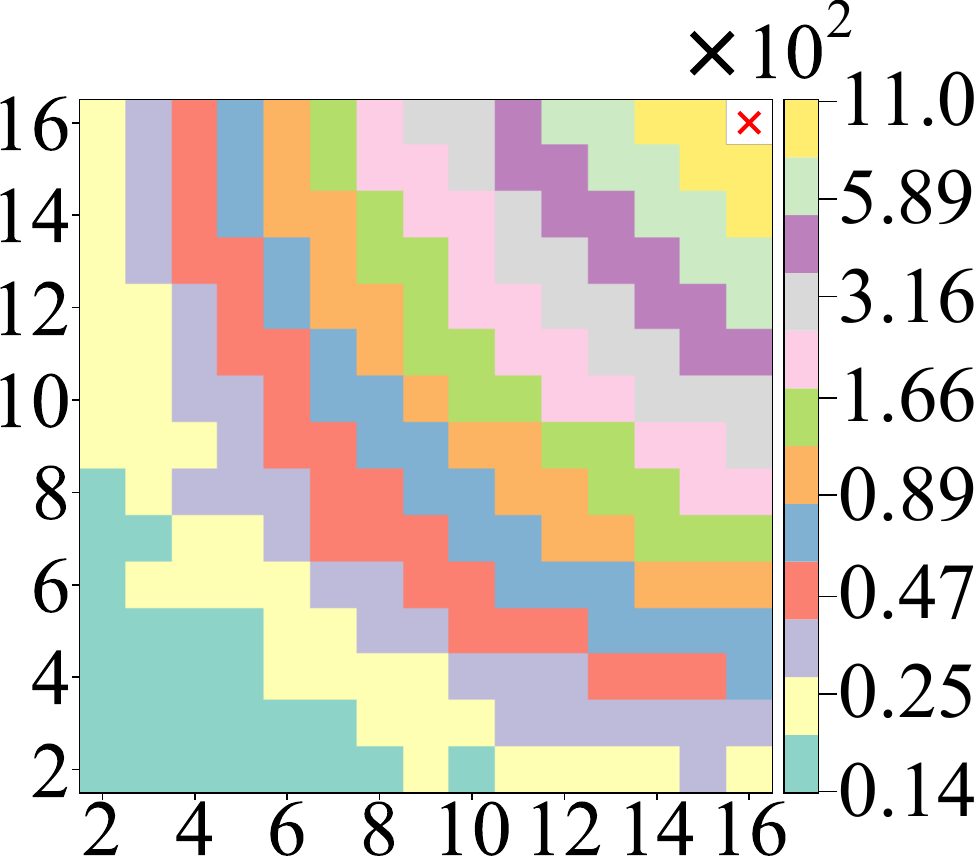}
        \caption{Atax}
        \label{fig:heatmap:atax}
    \end{subfigure}
    \hfill
    \begin{subfigure}[b]{0.16\textwidth}
        \includegraphics[width=\textwidth]{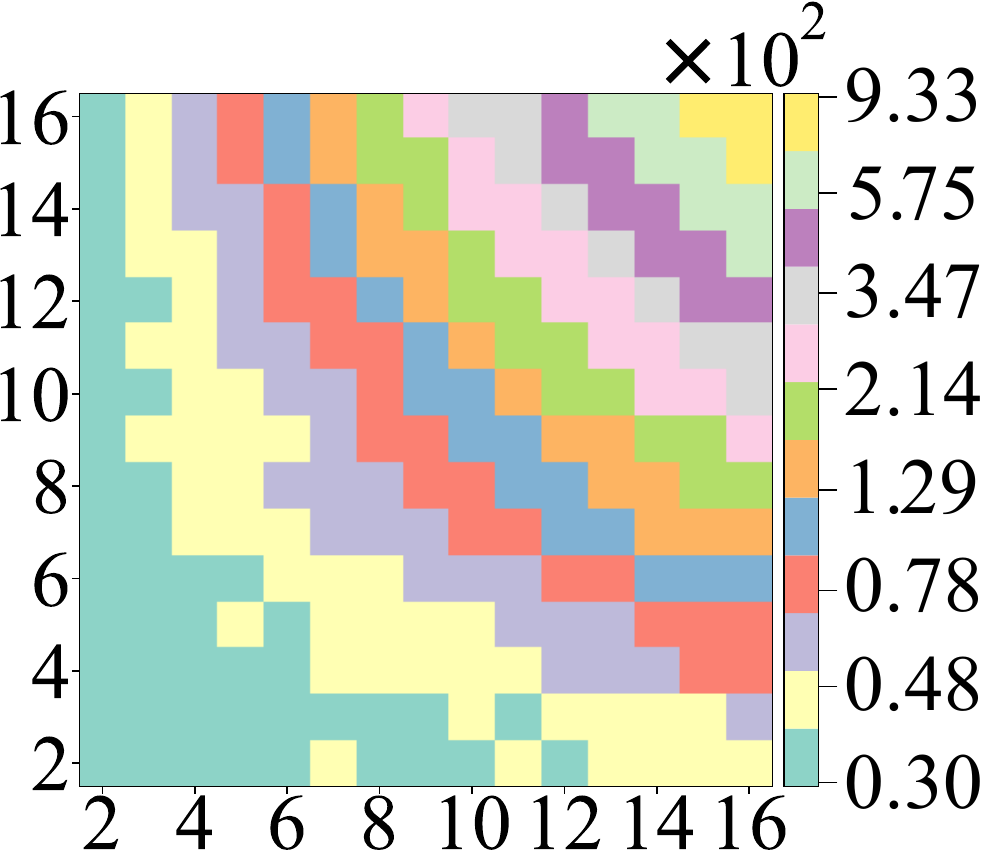}
        \caption{Bicg}
        \label{fig:heatmap:bicg}
    \end{subfigure}
    \begin{subfigure}[b]{0.16\textwidth}
        \includegraphics[width=\textwidth]{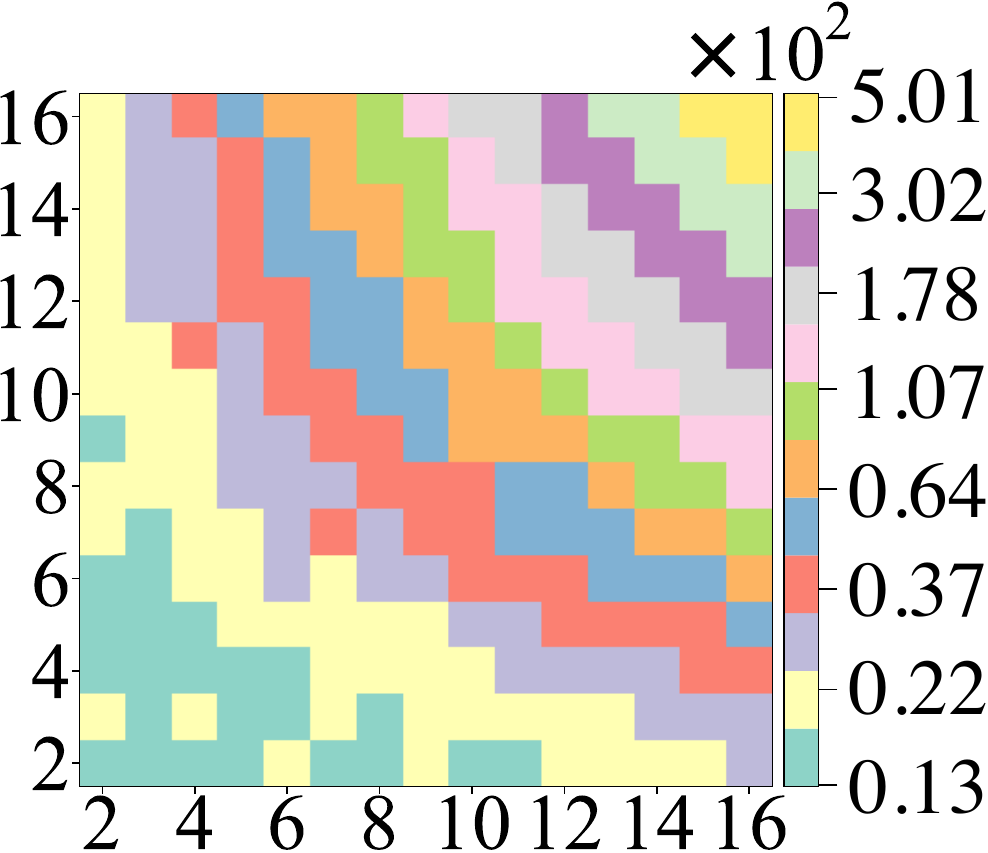}
        \caption{Gemm}
        \label{fig:heatmap:gemm}
    \end{subfigure}
    
    \medskip
    
    \begin{subfigure}[b]{0.16\textwidth}
        \includegraphics[width=\textwidth]{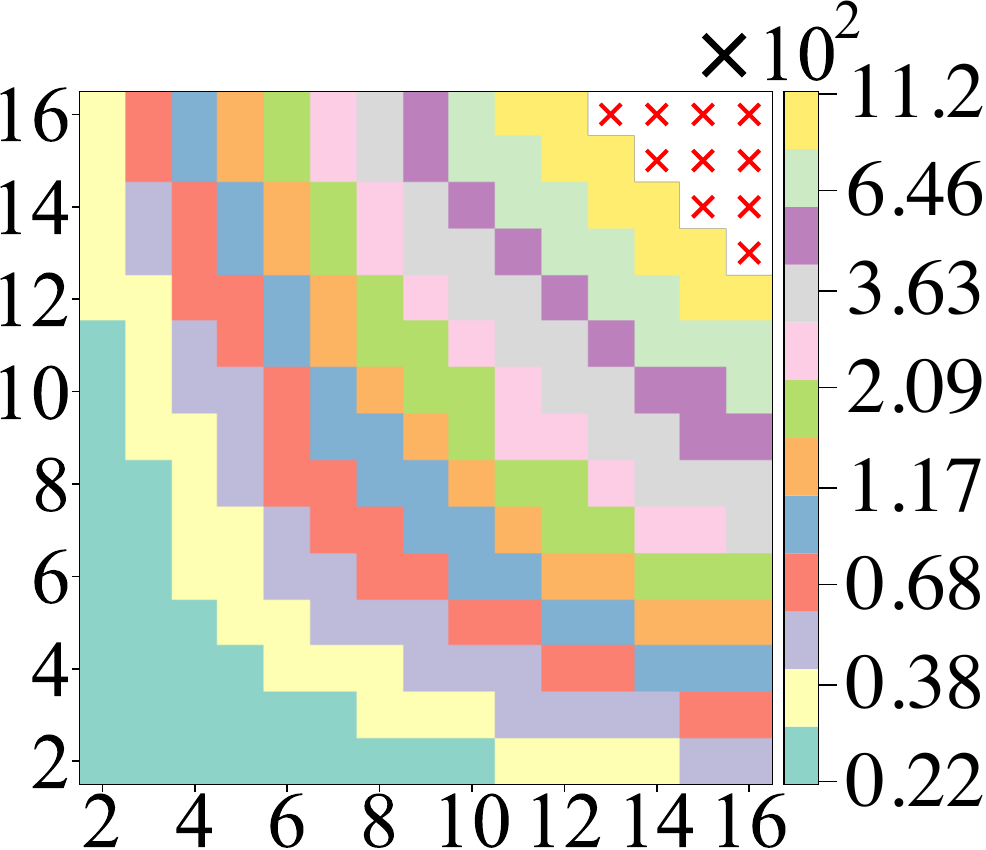}
        \caption{Seidel\_2d}
        \label{fig:seidel_2d}
    \end{subfigure}
    \hfill
    \begin{subfigure}[b]{0.16\textwidth}
        \includegraphics[width=\textwidth]{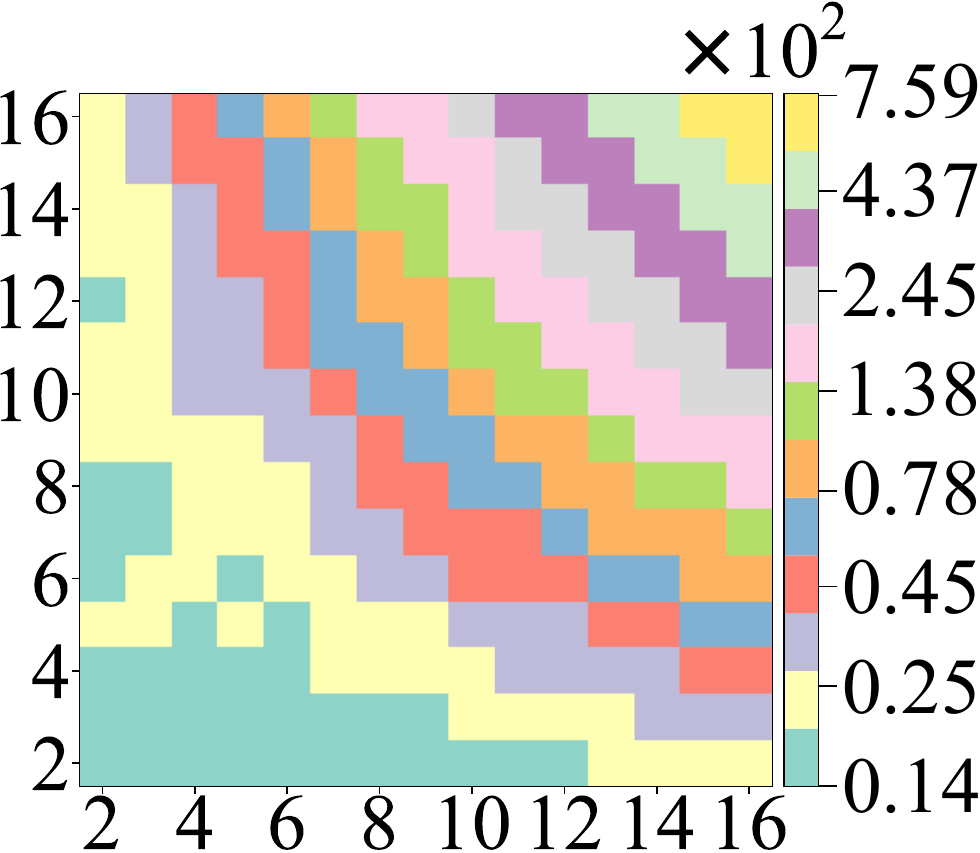}
        \caption{Mvt}
        \label{fig:heatmap:mvt}
    \end{subfigure}
    \hfill
    \begin{subfigure}[b]{0.16\textwidth}
        \includegraphics[width=\textwidth]{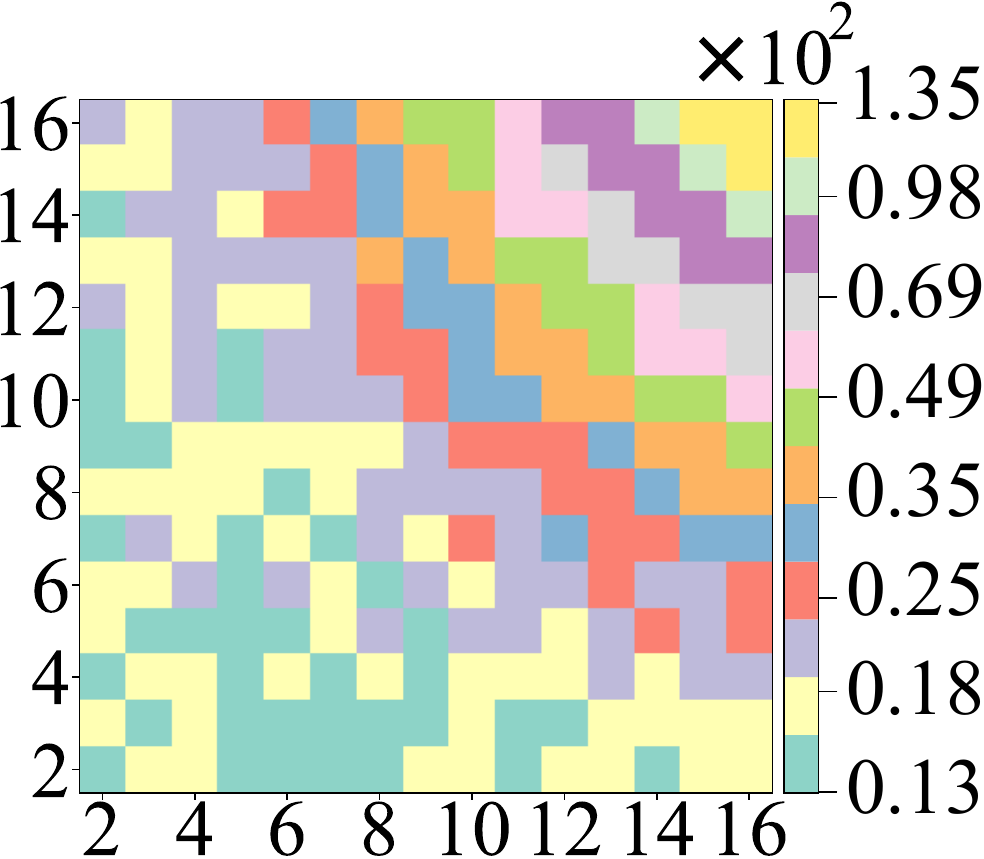}
        \caption{Trisolv}
        \label{fig:heatmap:trisolv}
    \end{subfigure}
    \hfill
    \begin{subfigure}[b]{0.16\textwidth}
        \includegraphics[width=\textwidth]{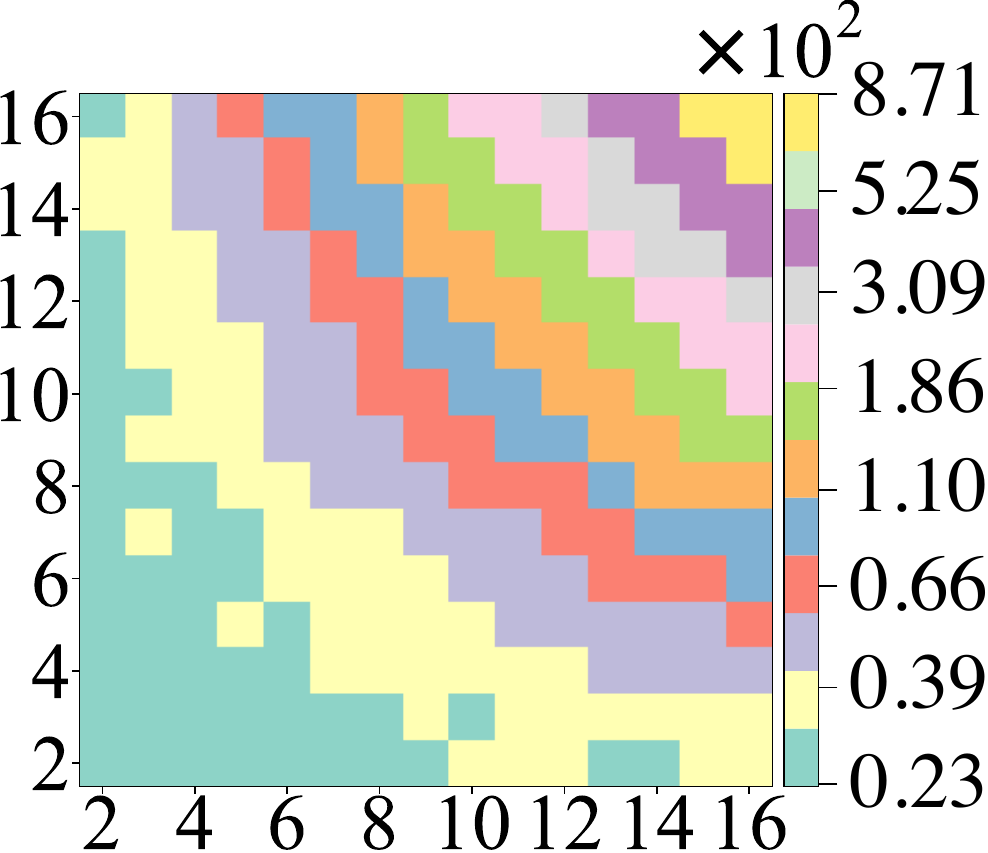}
        \caption{Gesummv}
        \label{fig:gesummv}
    \end{subfigure}
    \hfill
    \begin{subfigure}[b]{0.16\textwidth}
        \includegraphics[width=\textwidth]{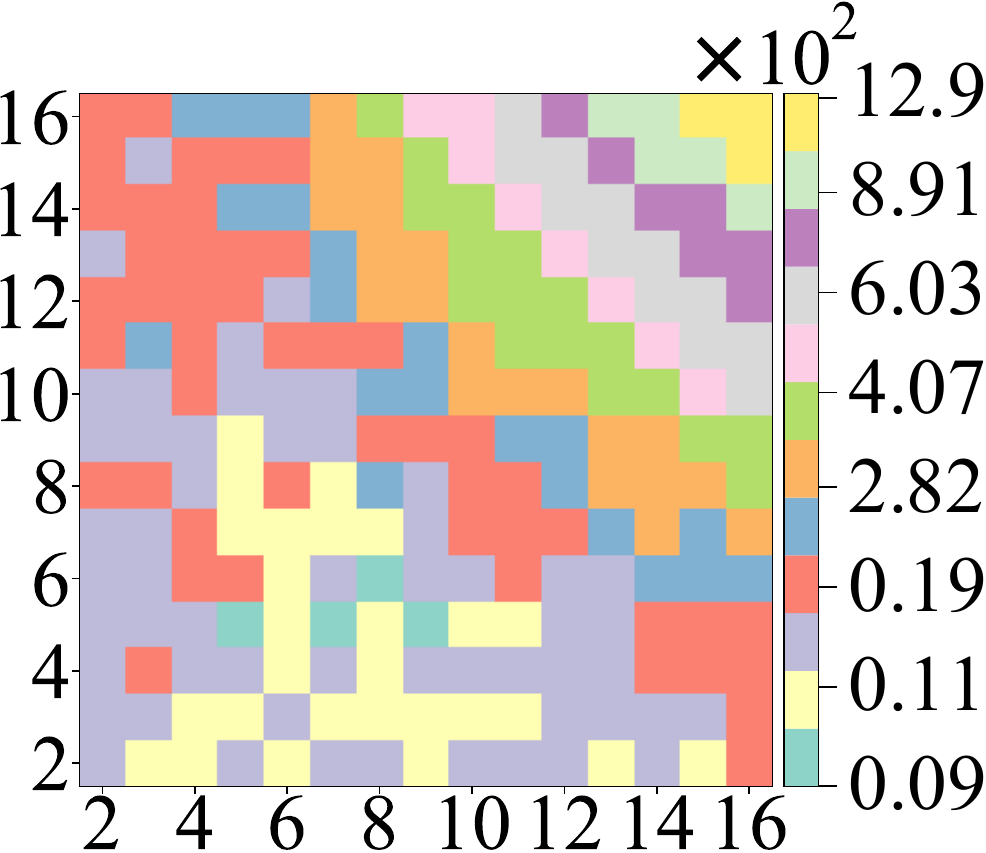}
        \caption{Trmm}
        \label{fig:heatmap:trmm}
    \end{subfigure}
    \hfill
    \begin{subfigure}[b]{0.16\textwidth}
        \includegraphics[width=\textwidth]{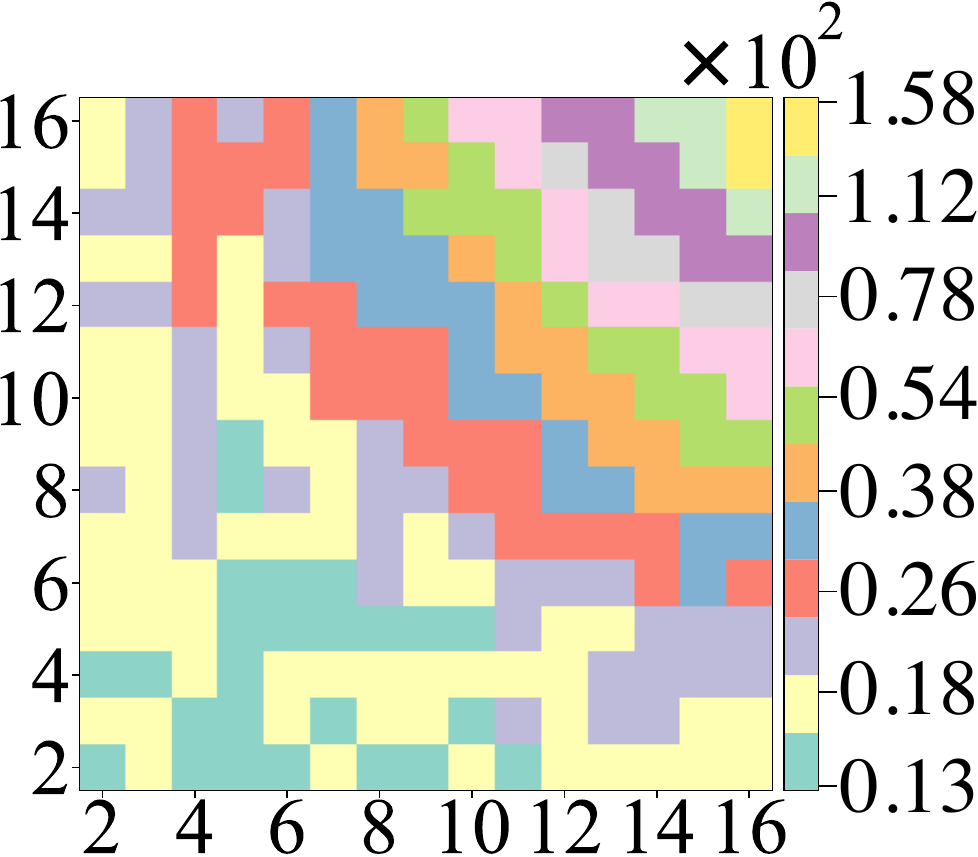}
        \caption{Cnn\_forward}
        \label{fig:heatmap:cnn_forward}
    \end{subfigure}
    \caption{Heatmap of Runtime for Different Nested Unrolling Configurations. The x-axis and y-axis represent the unrolling factors, while the colorbar on the right illustrates the end-to-end verification runtime (seconds).}
    \vspace{-3mm}
    \label{fig:heatmap}
\end{figure*}

The experiments were conducted on a system equipped with an Intel(R) Xeon(R) Gold 6418H CPU. This CPU has 48 physical cores, operates at a maximum speed of 4000 MHz, and the system has 1024 GB of RAM. We implemented transformations on selected benchmarks using the official MLIR compiler, \texttt{mlir-opt}, with LLVM version 18.0.0. 

There are many existing transformation validation tools in the literature, including CompCert\cite{leroy2009formal,leroy2016compcert}, Alive/Alive2\cite{lopes2021alive2,lopes2015provably}, PolyCheck\cite{bao2016polycheck}, and MLIR-TV\cite{bang2022smt}, each providing verification capabilities at different levels of the compilation stack. CompCert focuses on C-level compiler correctness, Alive/Alive2 target LLVM IR optimizations, PolyCheck handles affine code transformations for C-level compilation, while MLIR-TV is designed for MLIR-level verification. Among these tools, only MLIR-TV provides verification at the MLIR level, which is most relevant to our work. However, according to our testing, MLIR-TV fails to support the affine dialect and lacks coverage for affine operations that are central to high-level synthesis and the transformations evaluated in our benchmarks. This limitation renders a direct comparison with \texttt{\name} is challenging.

\begin{table}[!ht]
    \caption{List of benchmarks.}
    \label{tbl:benchmark}
    \small
    \centering
    \begin{tabular}{p{2.35cm}|p{3.7cm}|p{1.25cm}}
    \hline
        Benchmark & Description & Complexity \\ \hline
        GEMM~\cite{pouchet2012polybench} & General Matrix Multiply & \(O(n^3)\) \\ \hline
        LU~\cite{pouchet2012polybench} & LU Decomposition & \(O(n^3)\) \\ \hline
        2MM~\cite{pouchet2012polybench} & Two Matrix Multiplications & \(O(n^3)\) \\ \hline
        ATAX~\cite{pouchet2012polybench} & Matrix Transpose \newline Vector Multiplication & \(O(n^2)\) \\ \hline
        BiCG~\cite{pouchet2012polybench} & Biconjugate Gradient Method & \(O(n^2)\)\\ \hline
        GESUMMV~\cite{pouchet2012polybench} & SUM of \newline Matrix Vector Multiplications & \(O(n^2)\)\\ \hline
        MVT~\cite{pouchet2012polybench} & Matrix Vector Transpose & \(O(n^2)\)\\ \hline
        TRISOLV~\cite{pouchet2012polybench} & Triangular Solver & \(O(n^2)\)\\ \hline
        TRMM~\cite{pouchet2012polybench} & Triangular Matrix Multiply & \(O(n^3)\)\\ \hline
        CNN\_Forward~\cite{hrishikesh2018polybenchnn} & CNN Forward Function & \(O(n^7)\)\\ \hline
        Jacobi 1D~\cite{pouchet2012polybench} & Jacobi 1D iterative method & \makecell[l]{\(O(n \cdot t)\)}\\ \hline
        Seidel 2D~\cite{pouchet2012polybench} & Gauss-Seidel method & \makecell[l]{\(O(n^2 \cdot t)\)}\\ \hline
    \end{tabular}
    \vspace{-3mm}
\end{table}


\vspace{-3mm}
\subsection{Control Flow Transformation Verification}

\begin{table*}[!ht]
    \small    
    \caption{Runtime, dynamic rule numbers, e-classes numbers comparison under various tiling (T), unrolling setup (U) configurations. For benchmark Jacobi\_1d and Seidel\_2d, mlir-opt tool generates the Loop Boundary errors, which will be elaborated in Section \ref{sec:bugs}. Note that these are standard benchmarks used widely in MLIR-based compilation evaluations.}
    \vspace{4mm}
    \label{tbl:overall_label}
    \begin{adjustwidth}{-0.9cm}{-1.1cm}
    \centering
    \begin{tabular}{>{ \centering\arraybackslash}m{1.75cm}|>{ \centering\arraybackslash}m{2.08cm}|>{ \centering\arraybackslash}m{1.4cm}|>{ \centering\arraybackslash}m{0.6cm}|>{ \centering\arraybackslash}m{0.8cm}|>{ \centering\arraybackslash}m{0.6cm}|>{ \centering\arraybackslash}m{0.6cm}|>{ \centering\arraybackslash}m{0.6cm}|>{ \centering\arraybackslash}m{0.6cm}|>{ \centering\arraybackslash}m{0.85cm}|>{ \centering\arraybackslash}m{0.85cm}|>{ \centering\arraybackslash}m{0.85cm}|>{ \centering\arraybackslash}m{0.87cm}}
    \hline
         & Equivalence Checking & Metric & Base & T2\textasciitilde64 & U8 & U16 & U32 & U64 & T16-U8 & U16-T8 & U8-U4 & U16-U8 \\ \hline
        \multirow{4}{*}{2MM} & \multirow{4}{*}{\checkmark} & Runtime(s) & N/A & 7.3 & 6.8 & 7.8 & 10.9 & 25.3 & 6.8 & 7.9 & 27.8 & 173.8 \\ \cline{3-13}
         & & \#Dynamic Rules & N/A & 2  & 2  & 2  & 2  & 2  & 4  & 4  & 6  & 6 \\ \cline{3-13}
         & & \#E-classes & 156  & 198  & 427  & 683  & 1195  & 2219  & 469  & 725  & 1733  & 5069 \\ \hline
        \multirow{4}{*}{Jacobi\_1d} & \multirow{4}{*}{\color{red} \textbf{\makecell{Loop Boundary\\ Bug Identified}}} & Runtime(s) & N/A & 6.9 & 6.9 & 7.5 & 11.2 & 28.9 & 6.7 & 7.8 & 22.4 & 221.3 \\ \cline{3-13}
         & & \#Dynamic Rules & N/A & 1  & 2  & 2  & 2  & 2  & 3  & 3  & 6  & 6 \\ \cline{3-13}
         & & \#E-classes & 111  & 121  & 428  & 732  & 1340  & 2556  & 438  & 742  & 1832  & 5936 \\ \hline
        \multirow{4}{*}{Lu} & \multirow{4}{*}{\checkmark} & Runtime(s) & N/A & 6.9 & 6.7 & 6.9 & 9.6 & 21.4 & 6.7 & 7.3 & 23.5  & 155.1 \\ \cline{3-13}
         & & \#Dynamic Rules & N/A & 1  & 2  & 2  & 2  & 2  & 3  & 3  & 6  & 6 \\ \cline{3-13}
         & & \#E-classes & 110  & 119  & 364  & 604  & 1084  & 2044  & 373  & 613  & 1642  & 4718 \\ \hline
        \multirow{4}{*}{Atax} & \multirow{4}{*}{\checkmark} & Runtime(s) & N/A & 6.8 & 9.5 & 7.2 & 10.6 & 27.4 & 12.3 & 7.9 & 30.4 & 215.0 \\ \cline{3-13}
         & & \#Dynamic Rules & N/A & 2  & 3  & 3  & 3  & 3  & 6  & 4  & 9  & 9 \\ \cline{3-13}
         & & \#E-classes & 109  & 128  & 557  & 692  & 1252  & 2372  & 584  & 623  & 1930  & 5599 \\ \hline
        \multirow{4}{*}{Bicg} & \multirow{4}{*}{\checkmark} & Runtime(s) & N/A & 22.2 & 20.1 & 21.8 & 24.0 & 37.8 & 25.8 & 25.9 & 40.6 & 191.3 \\ \cline{3-13}
         & & \#Dynamic Rules & N/A & 2  & 2  & 2  & 2  & 2  & 5  & 3  & 6  & 6 \\ \cline{3-13}
         & & \#E-classes & 139  & 158  & 513  & 666  & 1178  & 2202  & 540  & 597  & 1735  & 5143 \\ \hline
        \multirow{4}{*}{Gemm} & \multirow{4}{*}{\checkmark} & Runtime(s) & N/A & 6.8 & 6.9 & 6.8 & 8.5 & 17.3 & 13.2 & 7.7 & 20.9 & 101.5 \\ \cline{3-13}
       & & \#Dynamic Rules & N/A & 1  & 2  & 2  & 2  & 2  & 3  & 3  & 6  & 6 \\ \cline{3-13}
         & & \#E-classes & 90  & 100  & 308  & 516  & 932  & 1764  & 318  & 526  & 1415  & 4078 \\ \hline
        \multirow{4}{*}{Seidel\_2d} & \multirow{4}{*}{\color{red} \textbf{\makecell{Loop Boundary\\ Bug Identified}}} & Runtime(s) & N/A & 14.5 & 15.8 & 16.7 & 22.0 & 51.0 & 14.7 & 13.3 & 36.5 & 380.1 \\ \cline{3-13}
         & & \#Dynamic Rules & N/A & 1  & 1  & 1  & 1  & 1  & 2  & 1  & 3  & 3 \\ \cline{3-13}
         & & \#E-classes & 214  & 246  & 563  & 907  & 1595  & 3017  & 584  & 246  & 2125  & 6723 \\ \hline
        \multirow{4}{*}{Mvt} & \multirow{4}{*}{\checkmark} & Runtime(s) & N/A & 6.8 & 6.3 & 6.9 & 9.2 & 20.6 & 6.1 & 6.0 & 25.7 & 144.7 \\ \cline{3-13}
         & & \#Dynamic Rules & N/A & 2  & 2  & 2  & 2  & 2  & 4  & 2  & 6  & 6 \\ \cline{3-13}
         & & \#E-classes & 91  & 131  & 339  & 579  & 1059  & 2019  & 356  & 131  & 1653  & 4683 \\ \hline
        \multirow{4}{*}{Trisolv} & \multirow{4}{*}{\checkmark} & Runtime(s) & N/A & 6.8 & 7.2 & 6.3 & 7.0 & 9.3 & 6.2 & 6.7 & 17.7 & 36.0 \\ \cline{3-13}
         & & \#Dynamic Rules & N/A & 1  & 1  & 1  & 1  & 1  & 2  & 2  & 3  & 3 \\ \cline{3-13}
         & & \#E-classes & 82  & 92  & 214  & 334  & 574  & 1054  & 224  & 344  & 900  & 2400 \\ \hline
        \multirow{4}{*}{Gesummv} & \multirow{4}{*}{\checkmark} & Runtime(s) & N/A & 19.5 & 19.1 & 19.3 & 21.4 & 31.1 & 17.8 & 18.3 & 30.3 & 131.5 \\ \cline{3-13}
         & & \#Dynamic Rules & N/A & 1  & 1  & 1  & 1  & 1  & 2  & 2  & 3  & 3 \\ \cline{3-13}
         & & \#E-classes & 172  & 181  & 402  & 618  & 1050  & 1914  & 411  & 627  & 1404  & 4372 \\ \hline
        \multirow{4}{*}{Trmm} & \multirow{4}{*}{\checkmark} & Runtime(s) & N/A & 6.7 & 5.8 & 6.2 & 6.6 & 9.2 & 6.2 & 8.4 & 16.4 & 31.6 \\ \cline{3-13}
         & & \#Dynamic Rules & N/A & 1  & 1  & 1  & 1  & 1  & 2  & 2  & 3  & 3 \\ \cline{3-13}
         & & \#E-classes & 72  & 93  & 201  & 321  & 561  & 1041  & 222  & 342  & 896  & 2381 \\ \hline
        \multirow{4}{*}{CNN\_forward} & \multirow{4}{*}{\checkmark} & Runtime(s) & N/A & 8.1 & 7.4 & 7.5 & 8.6 & 11.5 & 7.1 & 6.8 & 21.0 & 45.3 \\ \cline{3-13}
         & & \#Dynamic Rules & N/A & 1  & 1  & 1  & 1  & 1  & 2  & 1  & 3  & 3 \\ \cline{3-13}
         & & \#E-classes & 140  & 222  & 273  & 401  & 657  & 1169  & 343  & 222  & 983  & 2591 \\ \hline
    \end{tabular}
    \end{adjustwidth}
\end{table*}

We evaluate the performance of \name using the loop unrolling and tiling transformations on the MLIR benchmark. Specifically, we verify the equivalence between the input code within the selected benchmarks, reporting the end-to-end runtime, the number of dynamic rules implemented, and the number of e-classes. The benchmarks used in this evaluation are selected from Polybench \cite{pouchet2012polybench} and are enhanced with an additional benchmark, CNN-Forward, from Polybench-NN \cite{hrishikesh2018polybenchnn}. We utilize Polygeist \cite{moses2021polygeist} to convert the C code into MLIR, focusing specifically on the kernel portions for evaluation. These benchmarks are implemented as low-level kernels to facilitate hardware acceleration. The benchmarks, including description and time complexity, are shown in Table \ref{tbl:benchmark}.

Comprehensive results are provided in Table \ref{tbl:overall_label}, where all selected results have been successfully verified. The benchmarks for Seidel\_2d and Jacobi\_1d exhibit a specific type of error, classified as a \textbf{Loop Boundary Check Error}, which will be discussed in detail in Section \ref{sec:bugs}.

First, \name consistently verifies the correctness of transformations within one minute for most cases, except for nested unrolling. Second, the tiling transformation, which primarily adjusts loop parameters without altering code size, shows stable results across various tiling factors (represented as T2 to T64). Additionally, the number of e-classes has a significant impact on verification runtime, particularly in cases of nested unrolling. In these cases, the size of the e-graph predominantly determines the duration of verification.

\vspace{-3mm}

\subsection{Unrolling Verification Runtime}

\begin{figure*}[htb]
  \centering
  \begin{minipage}{1\linewidth}
    \centering
    \subcaptionbox{Runtime}
      {\includegraphics[width=0.535\linewidth]{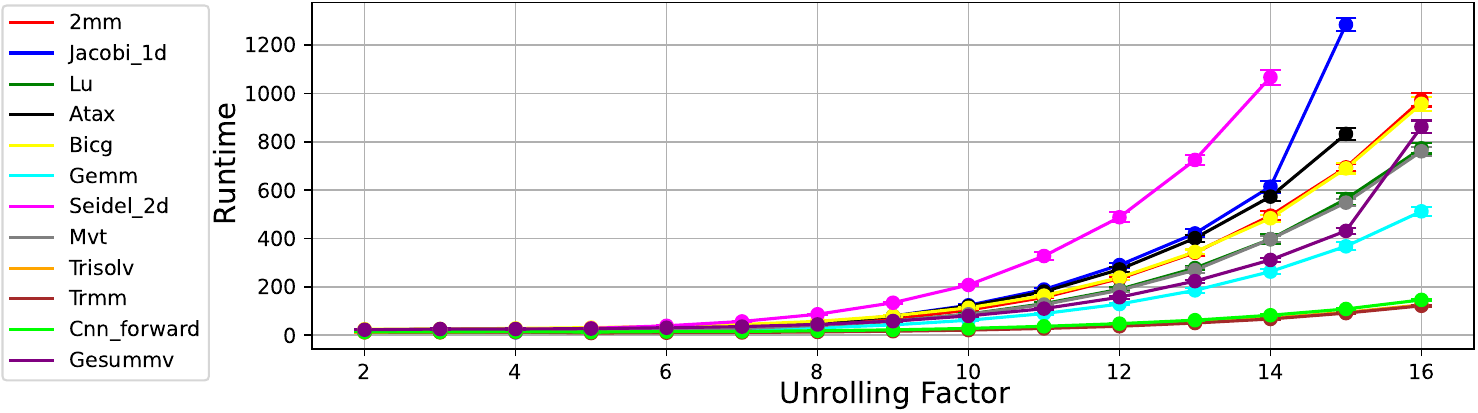}}
    \subcaptionbox{Number of e-class}
      {\includegraphics[width=0.460\linewidth]{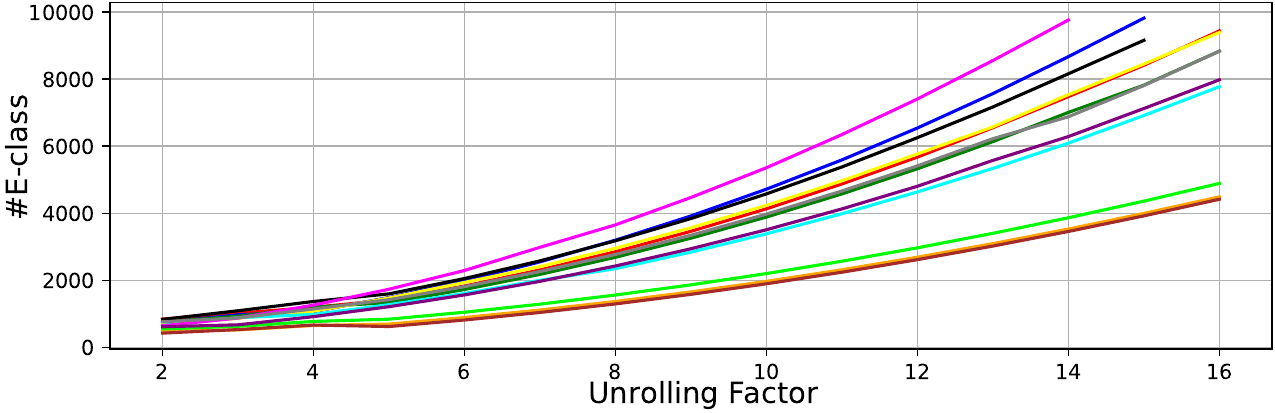}}
  \end{minipage}
  \vspace{-2mm}
  \caption{Runtime and e-graph complexity w.r.t nested unrolling complexity under the same unrolling factors.}
  \vspace{-5mm}
  \label{fig:runtime_eclass}
\end{figure*}

Since the unrolling transformation increases code size, unlike loop tiling, we specifically evaluate the verification runtime for nested unrolling as a scalability test. We assess the runtime performance of \name across various nested unrolling factors. The results are illustrated in Figure \ref{fig:heatmap}, where each sub-graph is a heatmap. The x and y axes of the heatmap represent unrolling factors from 2 to 16. The color of each pixel indicates the verification runtime by seconds, as detailed by the colorbar on the right. Cases marked with an "X" indicate that they either exceeded the time limit.

Figure \ref{fig:heatmap} shows that the verification runtime for most benchmarks completes within 20 minutes. E-graph saturation dominates the runtime, accounting for over 90\% in runtime. 

\vspace{9mm}

Additionally, we present detailed results in Figure \ref{fig:runtime_eclass} for the diagonal data sample from Figure \ref{fig:heatmap}. In this figure, the x-axis denotes the nested unrolling data sample (specifically, the unroll\_k\_unroll\_k, where k is the unrolling factor), and the y-axis displays the runtime and the number of e-classes.

This runtime is exponential rather than linear. Firstly, unrolling transformations modify code size, introducing scalability challenges, which are not present in other transformations like tiling and loop fusion. Secondly, the diagonal data sample involves nested unrolling, where the code size increases quadratically with each increase in the unrolling factor. This quadratic growth in code size significantly impacts runtime. Notably, it is uncommon for implementations to adopt such large unrolling factors due to the substantial increase in chip area it entails. From the perspective of power, performance, and area (PPA), such high unrolling factors are generally avoided, making the case of nested unrolling particularly rare.

\vspace{-3mm}

\subsection{Datapath Transformation Evaluation}
\vspace{-4mm}

\begin{figure}[!htb]
    \centering
    \includegraphics[width=0.9\linewidth]{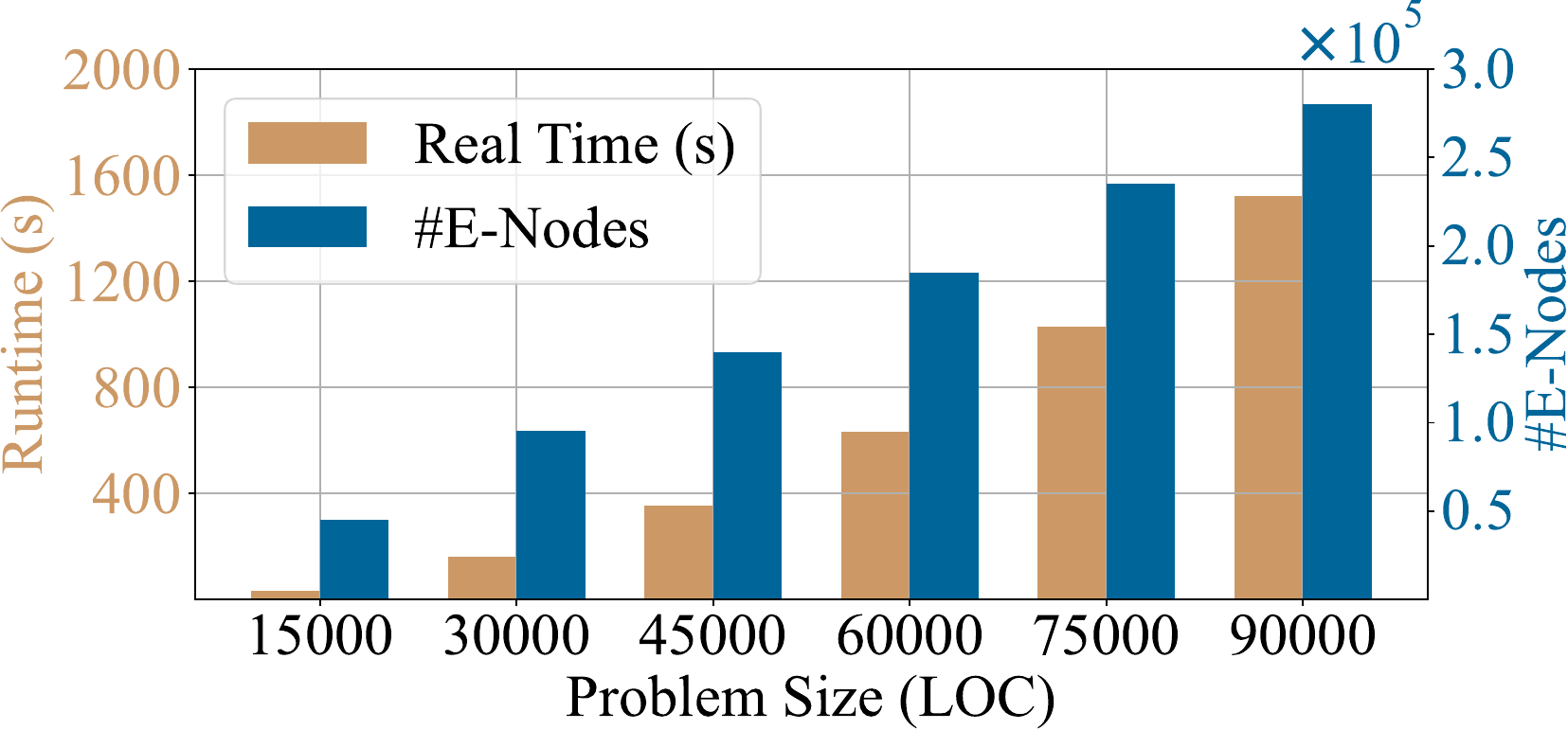}
    \vspace{-2mm}
    \caption{Runtime and e-nodes number for \name across representative problem sizes ranging from 15,000 to 90,000 LOC.}
    \label{fig:static}
    \vspace{-2mm}
\end{figure}

\name is designed to manage both static and dynamic transformations. In this section, we focus on the datapath transformation capabilities of \name. Specifically, we generated over 150 benchmarks incorporating various datapath transformations. These benchmarks vary in size, measured by lines of code (LOC), to assess the framework scalability for datapath transformation. We conducted experiments to measure both the runtime and the number of e-nodes generated for different benchmark sizes. The results are illustrated in Figure~\ref{fig:static}. In this figure, the x-axis represents the problem size measured in the sum of LOC for input benchmark, with representative datapoints ranging from 15,000 to 90,000, while the y-axis depicts the runtime (left) and the number of e-nodes (right).

Notably, all verification tasks were successfully completed within a 40-minute timeframe. For instance, the largest benchmark, encompassing 108,012 LOC, required a verification runtime of 2,305 seconds. This indicates that \name maintains consistent performance even as the problem size increases significantly. Additionally, the analysis of the number of e-nodes reveals a linear growth trend with respect to LOC, as illustrated in Figure~\ref{fig:static}. This linear relationship suggests that the complexity of the e-graph structure scales predictably with the size of the input code, ensuring manageable computational overhead. Together, these findings highlight the capability of \name to efficiently verify large-scale benchmarks while maintaining structural integrity and performance.

\vspace{-3mm}
\subsection{Mining Bugs in PolyBenchC MLIR}
\label{sec:bugs}

The MLIR affine dialect \footnote{\url{https://mlir.llvm.org/docs/Dialects/Affine/}} provides a suite of standard optimization transformations. We selected the representative kernels to showcase the application of these transformations. We will highlight two representative cases where the optimized HLS programs fail to preserve semantic equivalence with the original programs, and our tool has successfully detected and reported these discrepancies as bugs. In the following sections, we will elaborate on these two cases as case studies.





\textit{Case study 1: Loop Boundary Check Error:} We begin with an example in Listing \ref{lst:Case_study1_first} and its unrolled version (Listing \ref{lst:Case_study1_second}), which is unrolled by a factor of 2 using \texttt{mlir-opt}. The two code segments appear equivalent. The first loop in Listing \ref{lst:Case_study1_second} from lines 9 to 13 handles the majority of the operations, while the second loop from lines 16 to 18 processes the remainder. Together, both loops make up the entire code shown in Listing \ref{lst:Case_study1_first}. However, this transformation contains a flaw in the loop boundary checks. Specifically, if we consider the scenario where \texttt{\%0} equals 5, Listing \ref{lst:Case_study1_first} will not execute since the loop start value (15) exceeds its end value (10). In contrast, Listing \ref{lst:Case_study1_second} executes once for the second loop spanning from line 16 to line 19. This error occurs with all values of \texttt{\%0} less than 10.

Our tests indicate that these bugs are not confined to the specific case study presented. In general, for any loop defined with start and end conditions, \texttt{mlir-opt} preserves semantic equivalence only when the loop end value is guaranteed to exceed the loop start value. If this guarantee is not met and the loop is unrolled by a factor of \( k \), the transformed loop may inadvertently alter the loop boundaries, causing unintended iterations. For instance, such errors are observed in the Jacobi\_1d and Seidel\_2d benchmarks listed in Table~\ref{tbl:benchmark}.

\begin{figure}[t]
  \begin{minipage}{\columnwidth}
    \begin{lstlisting}[language=MLIR, caption=Case study 1, label=lst:Case_study1_first]
#map = affine_map<(d0) -> (d0 + 10)>
#map1 = affine_map<(d0) -> (d0 * 2)>
func.func @kernel(%arg0: i32, %arg1: memref<?xf64>) {
  %0 = arith.index_cast %arg0 : i32 to index
  // Loop range: from %0+10 to %0*2
  // Program should not execute when %0 is 5 (15->10)
  affine.for %arg2 = #map(%0) to #map1(%0) {
    %1 = affine.load %arg1[%arg2] : memref<?xf64>
  }
  return
}
    \end{lstlisting}
  \end{minipage}
  \vspace{-3mm}
  \begin{minipage}{\columnwidth}
    \begin{lstlisting}[language=MLIR, caption=Loop boundary check error: Loop unrolling, label=lst:Case_study1_second]
#map = affine_map<(d0) -> (d0 + 10)>
#map1 = affine_map<()[s0] -> (s0 + (s0 floordiv 2) * 2)>
#map2 = affine_map<(d0) -> (d0 + 1)>
#map3 = affine_map<(d0) -> (d0 * 2)>
func.func @kernel(%arg0: i32, %arg1: memref<?xf64>) {
  %0 = arith.index_cast %arg0 : i32 to index
  // Loop range: from %0+10 to %0+(%0 floordiv 2)*2
  // This loop should not execute when %0 is 5 (15->9)
  affine.for %arg2 = #map(%0) to #map1()[%0] step 2 {
    %1 = affine.load %arg1[%arg2] : memref<?xf64>
    %2 = affine.apply #map2(%arg2)
    %3 = affine.load %arg1[%2] : memref<?xf64>
  }
  // Loop range: from %0+(%0 floordiv 2)*2 to %0*2
  // This loop will execute once when %0 is 5 (9->10)
  affine.for %arg2 = #map1()[%0] to #map3(%0) {
    %1 = affine.load %arg1[%arg2] : memref<?xf64>
  }
  return
}
    \end{lstlisting}
  \end{minipage}
\end{figure}

\begin{figure}[t]
  \begin{minipage}{\columnwidth}
    \begin{lstlisting}[language=MLIR, caption=Case study 2, label=lst:Case_study2_first]
func.func @testing2(%arg0: memref<10xi32>, %arg1: memref<10xi32>) {
  %cst = arith.constant 1 : i32
  // Replace elements indexed from 1 to 10 with %arg0[0]
  affine.for %arg2 = 1 to 10 {
    %1 = affine.load %arg0[%arg2 - 1] : memref<10xi32>
    affine.store %1, %arg0[%arg2] : memref<10xi32>
  }
  // Elements indexed from 1 to 10 are incremented by 1
  affine.for %arg2 = 1 to 10 {
    %1 = affine.load %arg0[%arg2] : memref<10xi32>
    %2 = arith.addi %1, %cst : i32
    affine.store %2, %arg0[%arg2] : memref<10xi32>
  }
  // Finally, memory in %arg0 will be replaced to 
  // %arg0[0], %arg0[0]+1, %arg0[0]+1, ...
  return
}
    \end{lstlisting}
  \end{minipage}
  \vspace{-3mm}
  \begin{minipage}{\columnwidth}
    \begin{lstlisting}[language=MLIR, caption=Memory RAW violation: Loop fusion, label=lst:Case_study2_second]
func.func @testing2(%arg0: memref<10xi32>, %arg1: memref<10xi32>) {
  %cst = arith.constant 1 : i32
  affine.for %arg2 = 1 to 10 {
    // Load data from %arg0[%arg2 - 1]
    %0 = affine.load %arg0[%arg2 - 1] : memref<10xi32>
    affine.store %0, %arg0[%arg2] : memref<10xi32>
    %1 = affine.load %arg0[%arg2] : memref<10xi32>
    %2 = arith.addi %1, %cst : i32
    // Elements are incremented by 1
    affine.store %2, %arg0[%arg2] : memref<10xi32>
  }
  // Finally, memory in %arg0 will be replaced to 
  // %arg0[0], %arg0[0]+1, %arg0[0]+2, ...
  return
}
    \end{lstlisting}
  \end{minipage}
  \vspace{-6mm}
\end{figure}

\textit{Case study 2: Memory Read-After-Write Violation:} Listing \ref{lst:Case_study2_first} and Listing \ref{lst:Case_study2_second} illustrate another instance of semantic non-equivalence following a transformation—loop fusion. In the provided code, the divergence in memory state between the two functions arises from differences in the execution order of the loop body operations after loop fusion. In the first function, two separate loops handle the operations: the first loop copies values from one element to the next, and the second loop increments these values. As a result, the final state of the array consists of the original element \texttt{\%arg0[0]} followed by incremented versions of this element for all subsequent indices. Specifically, the array becomes \texttt{\%arg0[0]}, \texttt{\%arg0[0] + 1}, \texttt{\%arg0[0] + 1}, \dots, \texttt{\%arg0[0] + 1}, with every element after the first set to \texttt{\%arg0[0] + 1}.
However, in the fused version, each iteration of the single loop performs both copy and increment operations adjacently. This introduces a dependency violation on the increment operation of the previous iteration, leading to a linearly increasing sequence based on the first element. Finally, the array becomes \texttt{\%arg0[0]}, \texttt{\%arg0[0] + 1}, \texttt{\%arg0[0] + 2}, \dots, \texttt{\%arg0[0] + 10}. Such transformations in the loop structure affect not only the computational dependencies but also the overall memory state of \texttt{\%arg0}, resulting in non-equivalence of the final output across the two function versions.



In these case studies, we identify two types of transformation errors introduced by the \texttt{mlir-opt} compiler. Code unrolling leads to additional loop body executions, while loop fusion transformation changes the sequence of data dependencies, both affecting the intended behavior and memory state. Relying on \texttt{mlir-opt} for code transformations without subsequent equivalence checks can result in various adverse outcomes. Minor issues might include data corruption, while more severe implications could lead to complete system failures, potentially disrupting critical operations and compromising system integrity. Although code transformations can enhance performance, the improper application of these transformations without equivalence checking can introduce errors that disrupt control flow and alter semantic behavior.

\vspace{-3mm}
\section{Conclusions}

\label{sec:conclusion}

This paper introduces \name, a novel equivalence checking tool. \name leverages the dynamic rule generation to effectively verify complex control flow transformations. By integrating both static datapath rewriting rules with dynamic, code-specific rules, \name provides a flexible verification solution. Our experimental results demonstrate its efficiency, verifying various code transformations within an efficient runtime. Finally, \name successfully identifies a few critical compilation errors, covering functional incorrectness and data hazard violations. 

\textbf{Acknowledgment} -- This work is supported by National Science Foundation (NSF) under CCF2350186, CCF2403134, CCF2349670, and CCF2349461 awards and the U.S. Department of Energy, Office of Science, Office of Advanced Scientific Computing Research's Advanced Computing Technologies Competitive Portfolios program at Pacific Northwest National Laboratory (PNNL).

\bibliographystyle{plain}
\bibliography{reference,add}

\begin{thebibliography}{10}

\bibitem{The_IREE_Authors_IREE_2019}
{IREE}, September 2019.

\bibitem{isa}
{ISA}.
\newblock {\em \url{https://repo.or.cz/w/isa.git}}, 2024.

\bibitem{agostini2022soda}
Nicolas~Bohm Agostini, Serena Curzel, David Kaeli, and Antonino Tumeo.
\newblock Soda-opt an mlir based flow for co-design and high-level synthesis.
\newblock In {\em Proceedings of the 19th ACM International Conference on Computing Frontiers}, pages 201--202, 2022.

\bibitem{bachmair1994rewrite}
Leo Bachmair and Harald Ganzinger.
\newblock Rewrite-based equational theorem proving with selection and simplification.
\newblock {\em Journal of Logic and Computation}, 4(3):217--247, 1994.

\bibitem{bachmair1998equational}
Leo Bachmair and Harald Ganzinger.
\newblock Equational reasoning in saturation-based theorem proving.
\newblock {\em Automated deduction—a basis for applications}, 1:353--397, 1998.

\bibitem{bang2022smt}
Seongwon Bang, Seunghyeon Nam, Inwhan Chun, Ho~Young Jhoo, and Juneyoung Lee.
\newblock Smt-based translation validation for machine learning compiler.
\newblock In {\em International Conference on Computer Aided Verification}, pages 386--407. Springer, 2022.

\bibitem{bao2016polycheck}
Wenlei Bao, Sriram Krishnamoorthy, Louis-No{\"e}l Pouchet, Fabrice Rastello, and Ponnuswamy Sadayappan.
\newblock Polycheck: Dynamic verification of iteration space transformations on affine programs.
\newblock {\em ACM SIGPLAN Notices}, 51(1):539--554, 2016.

\bibitem{cao2023babble}
David Cao, Rose Kunkel, Chandrakana Nandi, Max Willsey, Zachary Tatlock, and Nadia Polikarpova.
\newblock Babble: Learning better abstractions with e-graphs and anti-unification.
\newblock {\em Proceedings of the ACM on Programming Languages}, 7(POPL):396--424, 2023.

\bibitem{chen2025morphic}
Chen Chen, Guangyu HU, Cunxi Yu, Yuzhe Ma, and Hongce Zhang.
\newblock E-morphic: Scalable equality saturation for structural exploration in logic\~{} synthesis.
\newblock {\em Design Automation Conference (DAC)}, 2025.

\bibitem{chen2024syn}
Chen Chen, Guangyu Hu, Dongsheng Zuo, Cunxi Yu, Yuzhe Ma, and Hongce Zhang.
\newblock E-syn: E-graph rewriting with technology-aware cost functions for logic synthesis.
\newblock pages 1--6, 2024.

\bibitem{cheng2023seer}
Jianyi Cheng, Samuel Coward, Lorenzo Chelini, Rafael Barbalho, and Theo Drane.
\newblock Seer: Super-optimization explorer for hls using e-graph rewriting with mlir.
\newblock {\em arXiv preprint arXiv:2308.07654}, 2023.

\bibitem{cheng2024seer}
Jianyi Cheng, Samuel Coward, Lorenzo Chelini, Rafael Barbalho, and Theo Drane.
\newblock Seer: Super-optimization explorer for high-level synthesis using e-graph rewriting.
\newblock In {\em Proceedings of the 29th ACM International Conference on Architectural Support for Programming Languages and Operating Systems, Volume 2}, pages 1029--1044, 2024.

\bibitem{circt}
Circuit ir compilers and tools ({CIRCT})., 2024.

\bibitem{cong2022fpga}
Jason Cong, Jason Lau, Gai Liu, Stephen Neuendorffer, Peichen Pan, Kees Vissers, and Zhiru Zhang.
\newblock Fpga hls today: successes, challenges, and opportunities.
\newblock {\em ACM Transactions on Reconfigurable Technology and Systems (TRETS)}, 15(4):1--42, 2022.

\bibitem{cong2011high}
Jason Cong, Bin Liu, Stephen Neuendorffer, Juanjo Noguera, Kees Vissers, and Zhiru Zhang.
\newblock High-level synthesis for fpgas: From prototyping to deployment.
\newblock {\em IEEE Transactions on Computer-Aided Design of Integrated Circuits and Systems}, 30(4):473--491, 2011.

\bibitem{coward2022automatic}
Samuel Coward, George~A Constantinides, and Theo Drane.
\newblock Automatic datapath optimization using e-graphs.
\newblock In {\em 2022 IEEE 29th Symposium on Computer Arithmetic (ARITH)}, pages 43--50. IEEE, 2022.

\bibitem{coward2023automating}
Samuel Coward, George~A Constantinides, and Theo Drane.
\newblock Automating constraint-aware datapath optimization using e-graphs.
\newblock In {\em 2023 60th ACM/IEEE Design Automation Conference (DAC)}, pages 1--6. IEEE, 2023.

\bibitem{coward2023datapath}
Samuel Coward, Emiliano Morini, Bryan Tan, Theo Drane, and George~A Constantinides.
\newblock Datapath verification via word-level e-graph rewriting.
\newblock In {\em 2023 Formal Methods in Computer-Aided Design (FMCAD)}, pages 92--100. IEEE, 2023.

\bibitem{dave2019dmazerunner}
Shail Dave, Youngbin Kim, Sasikanth Avancha, Kyoungwoo Lee, and Aviral Shrivastava.
\newblock Dmazerunner: Executing perfectly nested loops on dataflow accelerators.
\newblock {\em ACM Transactions on Embedded Computing Systems (TECS)}, 18(5s):1--27, 2019.

\bibitem{de2007efficient}
Leonardo De~Moura and Nikolaj Bj{\o}rner.
\newblock Efficient e-matching for smt solvers.
\newblock In {\em Automated Deduction--CADE-21: 21st International Conference on Automated Deduction Bremen, Germany, July 17-20, 2007 Proceedings 21}, pages 183--198. Springer, 2007.

\bibitem{de2008z3}
Leonardo De~Moura and Nikolaj Bj{\o}rner.
\newblock Z3: An efficient smt solver.
\newblock In {\em International conference on Tools and Algorithms for the Construction and Analysis of Systems}, pages 337--340. Springer, 2008.

\bibitem{dershowitz2005taste}
Nachum Dershowitz.
\newblock A taste of rewrite systems.
\newblock {\em Functional Programming, Concurrency, Simulation and Automated Reasoning: International Lecture Series 1991--1992 McMaster University, Hamilton, Ontario, Canada}, pages 199--228, 2005.

\bibitem{detlefs2005simplify}
David Detlefs, Greg Nelson, and James~B Saxe.
\newblock Simplify: a theorem prover for program checking.
\newblock {\em Journal of the ACM (JACM)}, 52(3):365--473, 2005.

\bibitem{jia2019taso}
Zhihao Jia, Oded Padon, James Thomas, Todd Warszawski, Matei Zaharia, and Alex Aiken.
\newblock Taso: optimizing deep learning computation with automatic generation of graph substitutions.
\newblock In {\em Proceedings of the 27th ACM Symposium on Operating Systems Principles}, pages 47--62, 2019.

\bibitem{karfa2013verification}
Chandan Karfa, Kunal Banerjee, Dipankar Sarkar, and Chittaranjan Mandal.
\newblock Verification of loop and arithmetic transformations of array-intensive behaviors.
\newblock {\em IEEE Transactions on Computer-Aided Design of Integrated Circuits and Systems}, 32(11):1787--1800, 2013.

\bibitem{lai2019heterocl}
Yi-Hsiang Lai, Yuze Chi, Yuwei Hu, Jie Wang, Cody~Hao Yu, Yuan Zhou, Jason Cong, and Zhiru Zhang.
\newblock Heterocl: A multi-paradigm programming infrastructure for software-defined reconfigurable computing.
\newblock In {\em Proceedings of the 2019 ACM/SIGDA International Symposium on Field-Programmable Gate Arrays}, pages 242--251, 2019.

\bibitem{lai2021programming}
Yi-Hsiang Lai, Ecenur Ustun, Shaojie Xiang, Zhenman Fang, Hongbo Rong, and Zhiru Zhang.
\newblock Programming and synthesis for software-defined fpga acceleration: status and future prospects.
\newblock {\em ACM Transactions on Reconfigurable Technology and Systems (TRETS)}, 14(4):1--39, 2021.

\bibitem{lattner2004llvm}
Chris Lattner and Vikram Adve.
\newblock {LLVM}: {A} compilation framework for lifelong program analysis \& transformation.
\newblock In {\em International symposium on code generation and optimization, 2004. CGO 2004.}, pages 75--86. IEEE, 2004.

\bibitem{lattner2021mlir}
Chris Lattner, Mehdi Amini, Uday Bondhugula, Albert Cohen, Andy Davis, Jacques Pienaar, River Riddle, Tatiana Shpeisman, Nicolas Vasilache, and Oleksandr Zinenko.
\newblock Mlir: Scaling compiler infrastructure for domain specific computation.
\newblock In {\em 2021 IEEE/ACM International Symposium on Code Generation and Optimization (CGO)}, pages 2--14. IEEE, 2021.

\bibitem{leroy2009formal}
Xavier Leroy.
\newblock Formal verification of a realistic compiler.
\newblock {\em Communications of the ACM}, 52(7):107--115, 2009.

\bibitem{leroy2016compcert}
Xavier Leroy, Sandrine Blazy, Daniel K{\"a}stner, Bernhard Schommer, Markus Pister, and Christian Ferdinand.
\newblock Compcert-a formally verified optimizing compiler.
\newblock In {\em ERTS 2016: Embedded Real Time Software and Systems, 8th European Congress}, 2016.

\bibitem{leverett1980overview}
Bruce~W Leverett, Roderic Geoffrey~Galton Cattell, Steven~O Hobbs, Joseph~M Newcomer, Andrew~Henry Reiner, Bruce~R Schatz, and William~A Wulf.
\newblock An overview of the production quality compiler-compiler project.
\newblock {\em Computer}, 13(8):38--49, 1980.

\bibitem{lopes2021alive2}
Nuno~P Lopes, Juneyoung Lee, Chung-Kil Hur, Zhengyang Liu, and John Regehr.
\newblock Alive2: bounded translation validation for llvm.
\newblock In {\em Proceedings of the 42nd ACM SIGPLAN International Conference on Programming Language Design and Implementation}, pages 65--79, 2021.

\bibitem{lopes2015provably}
Nuno~P Lopes, David Menendez, Santosh Nagarakatte, and John Regehr.
\newblock Provably correct peephole optimizations with alive.
\newblock In {\em Proceedings of the 36th ACM SIGPLAN Conference on Programming Language Design and Implementation}, pages 22--32, 2015.

\bibitem{moses2021polygeist}
William~S Moses, Lorenzo Chelini, Ruizhe Zhao, and Oleksandr Zinenko.
\newblock Polygeist: Raising c to polyhedral mlir.
\newblock In {\em 2021 30th International Conference on Parallel Architectures and Compilation Techniques (PACT)}, pages 45--59. IEEE, 2021.

\bibitem{nelson1980techniques}
Charles~Gregory Nelson.
\newblock {\em Techniques for program verification}.
\newblock Stanford University, 1980.

\bibitem{nelson1980fast}
Greg Nelson and Derek~C Oppen.
\newblock Fast decision procedures based on congruence closure.
\newblock {\em Journal of the ACM (JACM)}, 27(2):356--364, 1980.

\bibitem{nieuwenhuis2005proof}
Robert Nieuwenhuis and Albert Oliveras.
\newblock Proof-producing congruence closure.
\newblock In {\em International Conference on Rewriting Techniques and Applications}, pages 453--468. Springer, 2005.

\bibitem{panchekha2015automatically}
Pavel Panchekha, Alex Sanchez-Stern, James~R Wilcox, and Zachary Tatlock.
\newblock Automatically improving accuracy for floating point expressions.
\newblock {\em Acm Sigplan Notices}, 50(6):1--11, 2015.

\bibitem{pouchet2012polybench}
Louis-No{\"e}l Pouchet et~al.
\newblock Polybench: The polyhedral benchmark suite.
\newblock {\em \url{https://github.com/MatthiasJReisinger/PolyBenchC-4.2.1}}, 437:1--1, 2012.

\bibitem{pouchet2024formal}
Louis-No{\"e}l Pouchet, Emily Tucker, Niansong Zhang, Hongzheng Chen, Debjit Pal, Gabriel Rodr{\'\i}guez, and Zhiru Zhang.
\newblock Formal verification of source-to-source transformations for hls.
\newblock In {\em Proceedings of the 2024 ACM/SIGDA International Symposium on Field Programmable Gate Arrays}, pages 97--107, 2024.

\bibitem{ragan2013halide}
Jonathan Ragan-Kelley, Connelly Barnes, Andrew Adams, Sylvain Paris, Fr{\'e}do Durand, and Saman Amarasinghe.
\newblock Halide: a language and compiler for optimizing parallelism, locality, and recomputation in image processing pipelines.
\newblock {\em Acm Sigplan Notices}, 48(6):519--530, 2013.

\bibitem{shashidhar2005verification}
KC~Shashidhar, Maurice Bruynooghe, Francky Catthoor, and Gerda Janssens.
\newblock Verification of source code transformations by program equivalence checking.
\newblock In {\em Compiler Construction: 14th International Conference, CC 2005, Held as Part of the Joint European Conferences on Theory and Practice of Software, ETAPS 2005, Edinburgh, UK, April 4-8, 2005. Proceedings 14}, pages 221--236. Springer, 2005.

\bibitem{Smith_2021}
Gus~Henry Smith, Andrew Liu, Steven Lyubomirsky, Scott Davidson, Joseph McMahan, Michael Taylor, Luis Ceze, and Zachary Tatlock.
\newblock Pure tensor program rewriting via access patterns (representation pearl).
\newblock In {\em Proceedings of the 5th ACM SIGPLAN International Symposium on Machine Programming}, PLDI ’21. ACM, June 2021.

\bibitem{sohrabizadeh2022autodse}
Atefeh Sohrabizadeh, Cody~Hao Yu, Min Gao, and Jason Cong.
\newblock Autodse: Enabling software programmers to design efficient fpga accelerators.
\newblock {\em ACM Transactions on Design Automation of Electronic Systems (TODAES)}, 27(4):1--27, 2022.

\bibitem{tate2009equality}
Ross Tate, Michael Stepp, Zachary Tatlock, and Sorin Lerner.
\newblock Equality saturation: a new approach to optimization.
\newblock In {\em Proceedings of the 36th annual ACM SIGPLAN-SIGACT symposium on Principles of programming languages}, pages 264--276, 2009.

\bibitem{tillet2019triton}
Philippe Tillet, Hsiang-Tsung Kung, and David Cox.
\newblock Triton: an intermediate language and compiler for tiled neural network computations.
\newblock In {\em Proceedings of the 3rd ACM SIGPLAN International Workshop on Machine Learning and Programming Languages}, pages 10--19, 2019.

\bibitem{ustun2022impress}
Ecenur Ustun, Ismail San, Jiaqi Yin, Cunxi Yu, and Zhiru Zhang.
\newblock Impress: Large integer multiplication expression rewriting for fpga hls.
\newblock In {\em 2022 IEEE 30th Annual International Symposium on Field-Programmable Custom Computing Machines (FCCM)}, pages 1--10. IEEE, 2022.

\bibitem{ustun2023equality}
Ecenur Ustun, Cunxi Yu, and Zhiru Zhang.
\newblock Equality saturation for datapath synthesis: A pathway to pareto optimality.
\newblock In {\em 2023 60th ACM/IEEE Design Automation Conference (DAC)}, pages 1--2. IEEE, 2023.

\bibitem{hrishikesh2018polybenchnn}
Hrishikesh Vaidya, Akilesh B, Abhishek Patwardhan, and Ramakrishna Upadrasta.
\newblock Polybench-nn.
\newblock {\em \url{https://github.com/IITH-Compilers/PolyBench-NN}}, 2018.

\bibitem{vanhattum2021vectorization}
Alexa VanHattum, Rachit Nigam, Vincent~T Lee, James Bornholt, and Adrian Sampson.
\newblock Vectorization for digital signal processors via equality saturation.
\newblock In {\em Proceedings of the 26th ACM International Conference on Architectural Support for Programming Languages and Operating Systems}, pages 874--886, 2021.

\bibitem{verdoolaege2012equivalence}
Sven Verdoolaege, Gerda Janssens, and Maurice Bruynooghe.
\newblock Equivalence checking of static affine programs using widening to handle recurrences.
\newblock {\em ACM Transactions on Programming Languages and Systems (TOPLAS)}, 34(3):1--35, 2012.

\bibitem{willsey2021egg}
Max Willsey, Chandrakana Nandi, Yisu~Remy Wang, Oliver Flatt, Zachary Tatlock, and Pavel Panchekha.
\newblock Egg: Fast and extensible equality saturation.
\newblock {\em Proceedings of the ACM on Programming Languages}, 5(POPL):1--29, 2021.

\bibitem{woodruff2023rewriting}
Jackson Woodruff, Thomas Koehler, Alexander Brauckmann, Chris Cummins, Sam Ainsworth, and Michael~FP O'Boyle.
\newblock Rewriting history: Repurposing domain-specific cgras.
\newblock {\em arXiv preprint arXiv:2309.09112}, 2023.

\bibitem{wu2021ironman}
Nan Wu, Yuan Xie, and Cong Hao.
\newblock Ironman: Gnn-assisted design space exploration in high-level synthesis via reinforcement learning.
\newblock In {\em Proceedings of the 2021 on Great Lakes Symposium on VLSI}, pages 39--44, 2021.

\bibitem{yang2021equality}
Yichen Yang, Phitchaya Phothilimthana, Yisu Wang, Max Willsey, Sudip Roy, and Jacques Pienaar.
\newblock Equality saturation for tensor graph superoptimization.
\newblock {\em Proceedings of Machine Learning and Systems}, 3:255--268, 2021.

\bibitem{ye2022scalehls}
Hanchen Ye, Cong Hao, Jianyi Cheng, Hyunmin Jeong, Jack Huang, Stephen Neuendorffer, and Deming Chen.
\newblock Scalehls: A new scalable high-level synthesis framework on multi-level intermediate representation.
\newblock In {\em 2022 IEEE International Symposium on High-Performance Computer Architecture (HPCA)}, pages 741--755. IEEE, 2022.

\bibitem{yin2025boole}
Jiaqi Yin, Zhan Song, Chen Chen, Qihao Hu, and Cunxi Yu.
\newblock Boole: Exact symbolic reasoning via boolean equality saturation.
\newblock {\em Design Automation Conference (DAC)}, 2025.

\bibitem{yu2017advanced}
Cunxi Yu, Mihir Choudhury, Andrew Sullivan, and Maciej Ciesielski.
\newblock Advanced datapath synthesis using graph isomorphism.
\newblock In {\em 2017 IEEE/ACM International Conference on Computer-Aided Design (ICCAD)}, pages 424--429. IEEE, 2017.

\bibitem{egglog}
Yihong Zhang, Yisu~Remy Wang, Oliver Flatt, David Cao, Philip Zucker, Eli Rosenthal, Zachary Tatlock, and Max Willsey.
\newblock Better together: Unifying datalog and equality saturation.
\newblock {\em Proc. ACM Program. Lang.}, 7(PLDI), jun 2023.

\bibitem{zhao2022polsca}
Ruizhe Zhao, Jianyi Cheng, Wayne Luk, and George~A Constantinides.
\newblock Polsca: Polyhedral high-level synthesis with compiler transformations.
\newblock In {\em 2022 32nd International Conference on Field-Programmable Logic and Applications (FPL)}, pages 235--242. IEEE, 2022.

\end{thebibliography}

\end{document}